\documentclass[english]{achemso}
\usepackage[LGR,T1]{fontenc}
\usepackage[latin9]{inputenc}
\usepackage{verbatim}
\usepackage{textcomp}
\usepackage{bm}
\usepackage{amsmath}
\usepackage{graphicx}
\makeatletter

\title{Simulation of ab initio optical absorption spectrum of $\beta$-carotene
with fully resolved $S_{0}$ and $S_{2}$ vibrational normal modes}

\author{Mantas Jaku\v{c}ionis\textsuperscript{1}, Ignas Gaiži\={u}nas\textsuperscript{1},
Juozas Šulskus\textsuperscript{1}, Darius Abramavi\v{c}ius\textsuperscript{1}}

\affiliation{\textsuperscript{1}Institute of Chemical Physics, Vilnius University,
Sauletekio Ave. 9-III, LT-10222 Vilnius, Lithuania}

\DeclareRobustCommand{\greektext}{%
  \fontencoding{LGR}\selectfont\def\encodingdefault{LGR}}
\DeclareRobustCommand{\textgreek}[1]{\leavevmode{\greektext #1}}
\ProvideTextCommand{\~}{LGR}[1]{\char126#1}

\setkeys{acs}{maxauthors=10}
\setkeys{acs}{etalmode=truncate}
\email{darius.abramavicius@ff.vu.lt}

\makeatother

\usepackage{babel}
\begin{document}
\begin{abstract}
\label{abstract_label}Electronic absorption spectrum of $\beta$-carotene
($\beta$-Car) is studied using quantum chemistry and quantum dynamics
simulations. Vibrational normal modes were computed in optimized geometries
of the electronic ground state $S_{0}$ and the optically bright excited
$S_{2}$ state using the time-dependent density functional theory.
By expressing the $S_{2}$ state normal modes in terms of the ground
state modes, we find that no one-to-one correspondence between the
ground and excited state vibrational modes exists. Using the ab initio
results, we simulated $\beta$-Car absorption spectrum with all 282
vibrational modes in a model solvent at $300\ \text{K}$ using the
time-dependent Dirac-Frenkel variational principle (TDVP) and are
able to qualitatively reproduce the full absorption lineshape. By
comparing the 282-mode model with the prominent 2-mode model, widely
used to interpret carotenoid experiments, we find that the full 282-mode
model better describe the high frequency progression of carotenoid
absorption spectra, hence, vibrational modes become highly mixed during
the $S_{0}\rightarrow S_{2}$ optical excitation. The obtained results
suggest that electronic energy dissipation is mediated by numerous
vibrational modes.
\end{abstract}

\section{Introduction}

Pigment molecules in Nature form the basis of life on Earth by enabling
organisms to utilize the solar energy. Carotenoids form a unique class
of pigments with a conjugated polyene chain, responsible for light
absorption in green-blue color region. Over 700 carotenoid molecules
are found in Nature. They primarily play a role as coloring materials,
what underlie a vital and complex signalling processes \citep{Tian2015,Weaver2018}.
In photosynthesis carotenoids are essential in solar energy harvesting
and in photoprotection from oxygen damage. The latter emerge on microscopic
level, when light illumination is high, by formation of energy trapping
states \citep{Murchie2014,Ruban2016}. This trapping has been related
to quenching of the chlorophyll excited states by carotenoid singlet
state \citep{Ruban2007,Holt2005}, or by excitonic interaction between
chlorophyll and the carotenoid which is controlled by carotenoid conformations
\citep{Llansola-Portoles2017,Bode2009}. Carotenoids become thus responsible
for regulation of excitation energy fluxes in photosynthesis in volatile
conditions of daylight irradiation. One of the possible mechanisms
of such behavior involves a limited conformational rearrangement of
the protein scaffold, that could act as a molecular switch to activate
or deactivate the quenching mechanism \citep{Cupellini2020}. A strong
correlation between carotenoid and local environment deformations
is necessary for such mechanism to exist.

However, the primary deformations leading to carotenoid flexibility
are the molecular vibrations. They are usually induced during photon
absorption (and emission) and following excitation relaxation processes.
Probing excitation and vibration mediated relaxation processes in
carotenoids, necessary for understanding fundamental physical processes
involved in their functioning, is possible by performing time-resolved
optical spectroscopy experiments. It is well established that carotenoids
demonstrate a complex structure of electronic excited states \citep{Polivka2004,Llansola-Portoles2017a}
with at least three electronic states necessary to fully capture excitation
long-time dynamics. Direct optical excitation induces electronic $S_{0}\rightarrow S_{2}$
transition, where $S_{0}$ is the electronic ground state and $S_{2}$
is the first optically accessible (bright) electronic state, and the
optically dark electronic state $S_{1}$ lies between $S_{0}$ and
$S_{2}$. Additional intramolecular charge-transfer (CT) states have
been proposed in peridinin in agreement with the experimental results
\citep{Bautista1999,Frank2000}. Quantum chemical calculations using
the time-dependent density functional theory with the Tamm\textminus Dancoff
approximation \citep{Tamm1945,Dancoff1950,Andreussi2015} demonstrate
presence of CT state, which appears as the third and second excited
singlet state, respectively. Energy of the CT state has been shown
to decrease dramatically in solvents of increasing polarity, while
the energy of the dark $S_{1}$ state remains comparatively constant
\citep{Vaswani2003}. Several other types of electronic excited states
have been suggested, however, their existence and involvement in relaxation
process is still debatable \citep{Hashimoto2018}. Specific spectral
features have been assigned to $S_{1}$ and CT states and these may
play an important role in deexcitation processes \citep{Frank2000,Premvardhan2005,Cupellini2020}.

Vibrational heating and cooling is involved in the relaxation process
via the electonic-vibrational (vibronic) coupling \citep{BaleviciusJr2019a}.
Indeed, the strong vibronic coupling is rooted in a broad electronic
absorption spectra, more specifically, in a strong vibronic shoulder
for a range of different carotenoids as observed experimentally \citep{Mendes-Pinto2013}.
This feature is often associated with two vibrational modes: $\text{C-C}$
symmetric and asymmetric stretching vibrations with cumulative Huang-Rhys
factor larger than 1. These modes are known to be active in Raman
spectra and their frequencies scale linearly with the conjugation
length in carotenoids \citep{Llansola-Portoles2017a}. While molecular
vibrations affect symmetry properties of molecules, they do not affect
the oscillator strength of the dark state \citep{Wei2019}. Such empirical
effective 2-mode model has been extensively used for spectroscopy
simulations \citep{Polivka2001,Christensson2009,Balevicius2016,Fox2017,Gong2018,BaleviciusJr2019a}.
However, the two vibrational modes do not capture the high energy
vibrational wing and it is not clear whether the two modes are sufficient
to accurately describe the more complex ultrafast internal conversion
and energy transfer processes.

In this paper we present quantum chemistry and quantum dynamics description
of vibrational manifold of $\beta$-Car in its electronic states $S_{0}$
and $S_{2}$. We find that numerous vibrational modes become highly
mixed during the $S_{0}\rightarrow S_{2}$ optical excitation, resulting
in a complex $S_{2}$ state wavepacket. We are able to reveal the
full absorption spectrum, including the high-energy vibrational shoulder,
however, qualitatively correct vibrational peak ratios require the
raw quantum chemistry results to be scaled. Simulations thus suggest
that pathways responsible for the ultrafast electronic excitation
relaxation and internal conversion are mediated by numerous vibrational
modes resulting in a rapid and efficient electronic energy dissipation.

\section{Theoretical Methods\label{sec:Quantum-chemistry} \label{sec:-carotene-model}}

Quantum chemical analysis starts from the complete molecular Hamiltonian
including both electronic and vibrational degrees of freedom (DOFs)
\citep{Valkunasa,May2011a}. Using the Born-Oppenheimer approximation
the full Schrödinger equation is split into separate equations for
electronic and nuclear DOFs. The stationary Schrödinger equation for
electrons then parametrically depends on the nuclear coordinates
\begin{equation}
\hat{H}_{el}(\boldsymbol{R})\Phi_{m}(\boldsymbol{R})=E_{m}(\boldsymbol{R})\Phi_{m}(\boldsymbol{R}).
\end{equation}
 Here $\hat{H}_{el}$ includes electron kinetic energy, electronic
interaction with nuclei, electron-electron interactions and internuclear
interaction energy, $\boldsymbol{R}\equiv R_{1},R_{2}...$ labels
nuclei coordinates. Eigenvalues $E_{m}(\boldsymbol{R})$ and the corresponding
eigenstates $\Phi_{m}(\boldsymbol{R})$, which parametrically depend
on nuclei configuration, characterize electronic system.

Electronic energy minimum of the electronic ground state denotes the
reference point - the equilibrium molecular structure. If the nuclear
configuration deviates from the minimum, the electronic energy is
increased, hence, the electronic energy can be treated as the potential
energy for nuclei DOFs. For small deviations from the energy minimum,
we use the harmonic approximation, where the potential energy operator
is expanded up to quadratic terms. So, the potential energy for nuclei
displacements in electronic state $n$ can be written as (using Einstein
summation convention for repeating indices)
\begin{equation}
U_{n}(\boldsymbol{u})\approx\frac{1}{2}\mathcal{H}_{ij}^{(n)}u_{i}u_{j},
\end{equation}
 where we introduce mass-weighted Cartesian coordinates $u_{i}=\sqrt{M_{i}}\left(R_{i}-R_{i0}^{n}\right)$,
as the shifts of nuclei from their equilibrium positions, and
\begin{equation}
\mathcal{H}_{ij}^{(n)}=\frac{1}{\sqrt{M_{i}M_{j}}}\left.\frac{\partial^{2}E_{n}(\boldsymbol{R})}{\partial R_{i}\partial R_{j}}\right|_{\text{min}},
\end{equation}
is the Hessian matrix with derivatives taken at the global minimum
of the state $n$. Schrödinger equation for the nuclear wavefunctions
with respect to the specific electronic state $n$ is
\begin{equation}
\left(\hat{T}+\hat{U}_{n}(\boldsymbol{u})\right)\chi_{n\alpha}(\boldsymbol{u})=\varepsilon_{n\alpha}\chi_{n\alpha}(\boldsymbol{u}).\label{eq:vibr-Schroedinger}
\end{equation}
Here $\alpha$ is the vibrational quantum state index with energy
$\varepsilon_{n\alpha}$ and wavefunction $\chi_{\alpha n}$. Vibrational
Schrödinger equation splits into independent set of equations in the
normal coordinate representation; we denote these coordinates by $Q_{k}$.
Normal modes are obtained by diagonalizing the Hessian matrix for
each electronic state $n$. Solving the eigenstate equation
\begin{equation}
\mathcal{H}_{ij}^{(n)}L_{jk}^{(n)}=\omega_{nk}^{2}L_{ik}^{(n)},
\end{equation}
 yields normal mode frequencies $\omega_{nk}$, where $k$ labels
normal modes. The Hessian eigenvectors $L_{ik}^{(n)}$ relate normal
modes $k$ and nuclei displacements $u_{i}$. Placing eigenvectors
in columns, we form the matrix $L^{(n)}$, whose rank is $M=3N-6$
(six of the modes are physically irrelevant as 3 of them correspond
to the uniform translation of the whole molecule along Cartesian axes,
while the other 3 are uniform rotations about these axes, they are
excluded), and it is used to transform mass-weighted Cartesian internal
coordinates $u_{i}$ into normal coordinates $Q_{k}^{(n)}=\left(L^{(n)}\right)_{kl}^{-1}u_{l}$.

Complete description of the \emph{vibronic} molecular states, when
electronic and vibrational state is known, is given by the state vectors
$|n\bm{\alpha}_{n}\rangle$, where $\text{\ensuremath{\bm{\alpha}_{n}}\ensuremath{\equiv(\alpha_{n1},\alpha_{n2},...,\alpha_{nK})}}$
is the $M-$dimensional vector denoting vibrational states of all
vibrational modes in electronic state $n$. As normal modes are harmonic,
vibrational Hamiltonian in electronic state $n$ is given by 
\begin{equation}
\hat{H}^{(n)}=\frac{1}{2}\sum_{k}\left(\left(\hat{P}_{k}^{(n)}\right)^{2}+\omega_{nk}^{2}\left(\hat{Q}_{k}^{(n)}\right)^{2}\right)|n\rangle\langle n|.\label{eq:dim-Hamil}
\end{equation}

Absorption spectrum of a vibronic system involves all possible optical
transitions from the vibronic ground state $|g\bm{\beta}\rangle$
to the excited states $|e\bm{\alpha}\rangle$. Starting from the linear
response theory, the absorption spectrum is given by the Fourier transform
of the linear response function
\begin{equation}
S(\omega)=\frac{\omega}{nc}\text{Re}\int_{0}^{\infty}\text{d}te^{i\omega t}F\left(t\right),\label{eq:abs-fun}
\end{equation}
where n is the refraction index and c is the speed of light \citep{Mukamel1995,Valkunasa},
and
\begin{equation}
F\left(t\right)=\langle g\bm{\alpha}|e^{i\hat{H}_{g}t}\hat{\bm{P}}e^{-i\hat{H}_{e}t}\hat{\bm{P}}|g\bm{\alpha}\rangle,\label{eq:lin-resp-fun}
\end{equation}
is the dipole operator correlation function. In the Born approximation
the polarization operator $\hat{\bm{P}}$ acts only on electronic
DOFs, hence, $\hat{\boldsymbol{P}}=\bm{\mu}_{eg}^{(el)}\left(|e\rangle\langle g|+|g\rangle\langle e|\right)$
and $\bm{\mu}_{eg}^{(el)}$ is the electronic transition dipole. Matrix
elements of the polarization operator are given by
\begin{equation}
\langle e\bm{\alpha}|\hat{\bm{P}}|g\bm{\beta}\rangle=\bm{\mu}_{eg}^{(el)}\int\text{d}^{N}\boldsymbol{u}\prod_{j,k}\chi_{e\alpha_{j}}^{\star}(\boldsymbol{u})\chi_{g\beta_{k}}(\boldsymbol{u}).\label{eq:pol-oper-elements}
\end{equation}

The multi-dimensional integral correspond to vibrational overlaps
between vibrational wavefunction in different electronic states. Integral
computation is not trivial, because the sets of normal modes in different
electronic states are not orthogonal, transformation of one set of
normal modes into another is necessary \citep{Duschinsky1937,Sando2001,Meier2015}.

Difference of the set of normal modes in different electronic states
are characterized as follows. In electronic state $n$ the deviation
of atomic Cartesian coordinates $R_{i}$ from the equilibrium position
$R_{i0}^{(n)}$ may be expressed via the normal modes via relation
\begin{equation}
\sqrt{M_{i}}\left(R_{i}-R_{i0}^{(n)}\right)=L_{ij}^{(n)}Q_{j}^{(n)},
\end{equation}
and allow us to relate the relative mass-weighted atom shifts $D_{i}^{(mn)}\equiv\sqrt{M_{i}}\left(R_{i0}^{(m)}-R_{i0}^{(n)}\right)$
between the equilibrium positions in electronic states $m$ and $n$
as
\begin{equation}
D_{i}^{(mn)}=L_{ij}^{(n)}Q_{j}^{(n)}-L_{ij}^{(m)}Q_{j}^{(m)}.
\end{equation}
 Then the normal mode coordinates in state $m$ can be expressed in
terms of state $n$ normal mode coordinates as
\begin{equation}
Q_{i}^{(m)}=a_{ij}^{(mn)}Q_{j}^{(n)}-d_{i}^{(mn)},\label{eq:nm-expr}
\end{equation}
where the expansion coefficient of the $i$th normal mode in the $m$th
state in terms of the $j$th mode in $n$th state is
\begin{equation}
a_{ij}^{(mn)}\equiv\left(L^{(m)}\right)_{ik}^{-1}L_{kj}^{(n)},\label{eq:exp-coef}
\end{equation}
and the $i$th normal mode potential displacement in the $m$th state,
with respect to the position in the $n$th state, is
\begin{equation}
d_{i}^{(mn)}\equiv\left(L^{(m)}\right)_{ik}^{-1}D_{k}^{(mn)}.\label{eq:displacement}
\end{equation}
These are the two quantities that relate normal modes in different
electronic states. Likewise, normal mode momentum is also expanded
in terms of the $a_{kj}^{(eg)}$ coefficients (and zero displacement)
\begin{equation}
P_{i}^{(m)}\equiv a_{ij}^{(mn)}P_{j}^{(n)}.
\end{equation}

Further on we consider two electronic states: the ground state $|g\rangle$
and the electronic excited state $|e\rangle$. Instead of evaluating
propagators in Eq. (\ref{eq:lin-resp-fun}) by computing the multi-dimensional
vibrational overlaps in Eq. (\ref{eq:pol-oper-elements}), we choose
to specify a vibronic state basis using the coherent state representation,
and propagate it following the TDVP.

We begin with writing dimensionless Hamiltonian by introducing the
dimensionless momentum $\hat{p}_{k}^{(n)}\equiv\sqrt{\omega_{nk}}^{-1}\hat{P}_{k}^{(n)}$
and coordinate $\hat{q}_{k}^{(n)}\equiv\sqrt{\omega_{nk}}\hat{Q}_{k}^{(n)}$
operators for $n=g,e$. After inserting them in Eq. (\ref{eq:dim-Hamil}),
follows that the electronic ground state Hamiltonian is
\begin{equation}
\hat{H}^{(g)}=\sum_{k}\frac{\omega_{gk}}{2}\left(\left(\hat{p}_{k}^{(g)}\right)^{2}+\left(\hat{x}_{k}^{(g)}\right)^{2}\right)|g\rangle\langle g|,\label{eq:Hs_g}
\end{equation}
 and the electronic excited state $|e\rangle$ Hamiltonian is
\begin{align}
\hat{H}^{(e)} & =\left(\varepsilon_{e}+\Lambda_{e}^{\text{vib}}+\sum_{k}\frac{\omega_{ek}}{2}\left(\left(\hat{p}_{k}^{(e)}\right)^{2}+\left(\hat{x}_{k}^{(e)}\right)^{2}\right)\right)|e\rangle\langle e|\nonumber \\
 & -\sum_{k}\omega_{ek}\tilde{d}_{k}^{(eg)}\hat{x}_{k}^{(e)}|e\rangle\langle e|,\label{eq:Hs_e}
\end{align}
where $\tilde{d}_{k}^{(eg)}\equiv\sqrt{\omega_{ek}}d_{k}^{(eg)}$
is the dimensionless displacement and $\Lambda_{e}^{\text{vib}}\equiv\frac{1}{2}\sum_{k}\omega_{ek}\left(\tilde{d}_{k}^{(eg)}\right)^{2}$
is the total vibrational reorganization energy. The resulting operators
in Eqs. (\ref{eq:Hs_g}-\ref{eq:Hs_e}) read 
\begin{equation}
\hat{x}_{k}^{(n)}\equiv\beta_{nk,j}a_{kj}^{(ng)}\hat{q}_{j}^{(g)},\label{eq:x_op}
\end{equation}
\begin{equation}
\hat{p}_{k}^{(n)}\equiv\beta_{nk,j}^{-1}a_{kj}^{(ng)}\hat{p}_{j}^{(g)},\label{eq:p_op}
\end{equation}
where $\beta_{nk,j}\equiv\sqrt{\omega_{nk}/\omega_{gj}}$. Eq. (\ref{eq:x_op})
and (\ref{eq:p_op}) describe the dimensionless coordinate and momentum
of the $k$th normal mode about its equilibrium point in the $n$th
electronic state. Terms $\beta_{nk,j}$ appear due to the normal mode
mixing and different vibrational frequencies in state $g$ and $e$.
We also add $\varepsilon_{e}$ as the purely electronic excitation
energy and set $\varepsilon_{g}=0\ \text{cm}^{-1}$. The total system
Hamiltonian is the sum over all electronic state terms $\hat{H}_{S}=\hat{H}^{(g)}+\hat{H}^{(e)}$.

Solvent effects will be simulated by considering energy fluctuations
of the molecular environment. Thermal fluctuations are induced by
a set of the quantum harmonic oscillators of a given temperature,
we will refer to this subsystem as the phonon bath. The phonon bath
Hamiltonian is
\begin{equation}
\hat{H}_{\text{B}}=\sum_{p}\frac{w_{p}}{2}\left(\hat{\rho}_{p}^{2}+\hat{\chi}_{p}^{2}\right),
\end{equation}
where $w_{p}$ is the frequency of the $p$th phonon mode, while $\hat{\rho}_{p}$
and $\hat{\chi}_{p}$ are the momentum and the coordinate operators,
respectively. Interaction between the system electronic states and
the phonon bath is included using the displaced oscillator model \citep{May2011a},
with the system-bath interaction Hamiltonian
\begin{equation}
\hat{H}_{\text{S-B}}=-\sum_{p}w_{p}f_{ep}\hat{\chi}_{p}|e\rangle\langle e|,\label{eq:H_sb}
\end{equation}
here $f_{ep}$ is the electron-phonon coupling strength of the $p$th
phonon mode to the electronic state $e$. The electronic ground state
is taken as the reference point so it is not affected by bath fluctuations
$f_{gp}=0$. Notice, that the system-bath coupling has the same form
as the last term in Eq. (\ref{eq:Hs_e}). The electronic state energy
modulation by the intramolecular and intermolecular vibrations is
treated equivalently. Likewise, we get additional contribution to
the reorganization energy $\Lambda_{e}^{\mathrm{\text{ph}}}=\frac{1}{2}\sum_{p}w_{p}f_{ep}^{2}$.
Usually, all excited electronic states are described as having the
same coupling strength to the bath, thus, changing all states' energies
by the same amount. For simplicity, we absorb $\Lambda_{e}^{\mathrm{\text{ph}}}$
into the definition of the excited state energy $\varepsilon_{e}$,
however, $\Lambda_{e}^{\mathrm{\text{ph}}}$ is still used to define
the electron-phonon coupling strengths $f_{ep}$. The full model Hamiltonian
is the sum of terms
\begin{equation}
\hat{H}=\hat{H}_{S}+\hat{H}_{\text{B}}+\hat{H}_{\text{S-B}}.
\end{equation}

Fluctuation characteristics of the phonon bath can be represented
by the spectral density function
\begin{equation}
C_{e}^{"}\left(\omega\right)=\frac{\pi}{2}\sum_{p}f_{ep}^{2}w_{p}^{2}\left[\delta\left(\omega-w_{p}\right)-\delta\left(\omega+w_{p}\right)\right],\label{eq:C"}
\end{equation}
where $\delta\left(\omega\right)$ is the Dirac delta function. Integration
of Eq. (\ref{eq:C"}) over the complete frequency range defines the
phonon bath reorganization energy in the $n$th electronic state
\begin{equation}
\Lambda_{e}^{\text{ph}}=\frac{1}{\pi}\int_{0}^{\infty}\frac{C_{e}^{"}\left(w\right)}{w}\text{d}w=\frac{1}{2}\sum_{p}w_{p}f_{ep}^{2}.\label{eq:Lphon}
\end{equation}

Many theories have been proposed to evaluate the linear response function
and the necessary polarization operator matrix elements (Eqs. \ref{eq:lin-resp-fun},
\ref{eq:pol-oper-elements}). Notibly, the foundational theory by
Yan and Mukamel \citep{Yan1998}, Franck-Condon approaches \citep{Borrelli2003,Ianconescu2004,BorrelliRaffaele2013},
as well as, the theories that include include non-Condon effects \citep{Niu2010,Borrelli2012,Baiardi2013,Toutounji2020}.

We chose to compute the linear response function by propagating the
Davydov $\text{D}_{2}$ trial wavefunction originating from the molecular
chain soliton theory \citep{Davydov1979,Scott1991}. For $N$ electronic
states, we can write an arbitrary state of the system as a superposition
-- the Davydov $\text{D}_{2}$ wavefunction is 
\begin{align}
|\Psi\left(t\right)\rangle & =\sum_{n}\alpha_{n}(t)\underbrace{|n\rangle\times|\tilde{\lambda}_{1}\left(t\right),\tilde{\lambda}_{2}\left(t\right),\ldots,\tilde{\lambda}_{K}\left(t\right)\rangle}_{\text{molecule state}}\nonumber \\
 & \times\underbrace{|\lambda_{1}\left(t\right),\lambda_{2}\left(t\right),\ldots,\lambda_{P}\left(t\right)\rangle}_{\text{solvent phonon state}}.\label{eq:D2-1}
\end{align}
It utilises coherent state representation for all vibrational modes.
For the shifted harmonic oscillator model, coherent states results
in an exact dynamics \citep{Choi2004}. $\alpha_{n}\left(t\right)$
is the amplitude of electronic state $|n\rangle$, in our case $n=g,e$.
Vibrational and phonon bath modes are represented using coherent states
$|\lambda\left(t\right)\rangle=\exp\left(\lambda\left(t\right)\hat{b}^{\dagger}-\lambda^{\star}\left(t\right)\hat{b}\right)|0\rangle$,
defined with respect to electronic ground state vibrational modes,
where $\lambda\left(t\right)$ is the coherent state parameter, and
$|0\rangle$ is the vacuum state of a quantum harmonic oscillator.
$\hat{b}_{i}^{\dagger}$ and $\hat{b}_{i}$ are the corresponding
bosonic creation and annihilation operators. Only the electronic ground
state normal modes are represented by the coherent states, modes of
the excited state are expanded in terms of the ground state coherent
states. Davydov type wavefunctions have been extensively used to model
single molecule, as well as, their aggregate dynamics \citep{Sun2010b,Chorosajev2016a,Wang2016b,Jakucionis2018c,Jakucionis2020a},
linear and nonlinear spetra \citep{Sun2015a,Zhou2016,Chorosajev2017a,Somoza2017a,Chen2019a}.

Time evolution of the Davydov $\text{D}_{2}$ wavefunction is obtained
by applying the Euler-Lagrange equation
\begin{equation}
\frac{\text{d}}{\text{d}t}\left(\frac{\partial\mathcal{L}\left(t\right)}{\partial\dot{\eta}_{i}^{\star}\left(t\right)}\right)-\frac{\partial\mathcal{L}\left(t\right)}{\partial\eta_{i}^{\star}\left(t\right)}=0,\label{eq:euler-lagrange}
\end{equation}
to each of the time-dependent parameter $\eta_{i}=\alpha_{n},\tilde{\lambda}_{k},\lambda_{p}$,
where
\begin{equation}
\mathcal{L}\left(t\right)=\frac{\text{i}}{2}\left(\langle\Psi|\frac{\text{d}}{\text{d}t}\Psi\rangle-\langle\frac{\text{d}}{\text{d}t}\Psi|\Psi\rangle\right)-\langle\Psi|\hat{H}|\Psi\rangle,\label{eq:Lagrangian}
\end{equation}
is the Lagrangian of the model given in terms of the Hamiltonian $\hat{H}$.
For convenience, we ommit explicitly writing time dependence. Euler-Lagrange
equation yields a system of coupled differential equations for the
$\alpha_{n}$, $\tilde{\lambda}_{k}$, $\lambda_{p}$ parameters of
the Davydov $\text{D}_{2}$ wavefunction, see Supplementary Information
for the full derivation. Equations describing model dynamics while
system is in the excited state $|e\rangle$ are
\begin{align}
\frac{\text{d}}{\text{d}t}\alpha_{e}= & -\text{i}\alpha_{e}\left(\varepsilon_{e}+\Lambda_{e}^{\text{vib}}+\sum_{k,j}\frac{\omega_{ek}^{2}+\omega_{gj}^{2}}{4\omega_{gj}}\left(a_{kj}^{(eg)}\right)^{2}\right)\nonumber \\
 & +\text{i}\alpha_{e}\left(\sum_{k}\frac{\omega_{ek}}{2}d_{k}^{(eg)}x_{k}^{(e)}+\sum_{q}w_{q}\frac{f_{eq}}{\sqrt{2}}\text{Re}\lambda_{q}\right),\label{eq:a_e}
\end{align}
\begin{align}
\frac{\text{d}}{\text{d}t}\tilde{\lambda}_{k}= & -\text{i}\sum_{j}\frac{\omega_{ej}}{\sqrt{2}}\beta_{ej,k}a_{jk}^{(eg)}\left(x_{j}^{(e)}-d_{j}^{(eg)}+\text{i}p_{j}^{(e)}\right),\label{eq:il_e}
\end{align}
\begin{align}
\frac{\text{d}}{\text{d}t}\lambda_{q}= & -\text{i}w_{q}\left(\lambda_{q}-\frac{f_{eq}}{\sqrt{2}}\right).\label{eq:l_e}
\end{align}
where $x_{k}^{(e)}=\beta_{ek,j}a_{kj}^{(eg)}\sqrt{2}\text{Re}\tilde{\lambda}_{j}$
and $p_{k}^{(e)}=\beta_{ek,j}^{-1}a_{kj}^{(eg)}\sqrt{2}\text{Im}\tilde{\lambda}_{j}$
are the expectation values of operators in Eq. (\ref{eq:x_op}) and
(\ref{eq:p_op}). The resulting system of equations for the ground
state $|g\rangle$ dynamics can be solved analytically $\alpha_{g}\left(t\right)=\alpha_{g}\left(0\right)\exp\left(-\frac{\text{i}}{2}\sum_{k}\omega_{gk}t\right)$,
$\tilde{\lambda}_{k}\left(t\right)=\tilde{\lambda}_{k}\left(0\right)\exp\left(-\text{i}\omega_{gk}t\right),$
$\lambda_{q}\left(t\right)=\lambda_{q}\left(0\right)\exp\left(-\text{i}w_{q}t\right)$.
Separation of equations into the ground and the excited state manifold
is convenient for the computation of the optical observables using
the response function theory. Terms due to the mixing of normal modes
remain present in Eq. (\ref{eq:a_e}) and (\ref{eq:il_e}). In the
latter, evolution of the $k$th mode is influenced by the motion of
all other $j$th modes. Eqs. (\ref{eq:a_e}-\ref{eq:l_e}) were solved
numerically using the adaptive step size Runge-Kutta algorithm.
\begin{figure*}[!t]
\centering{}\includegraphics[width=0.85\textwidth]{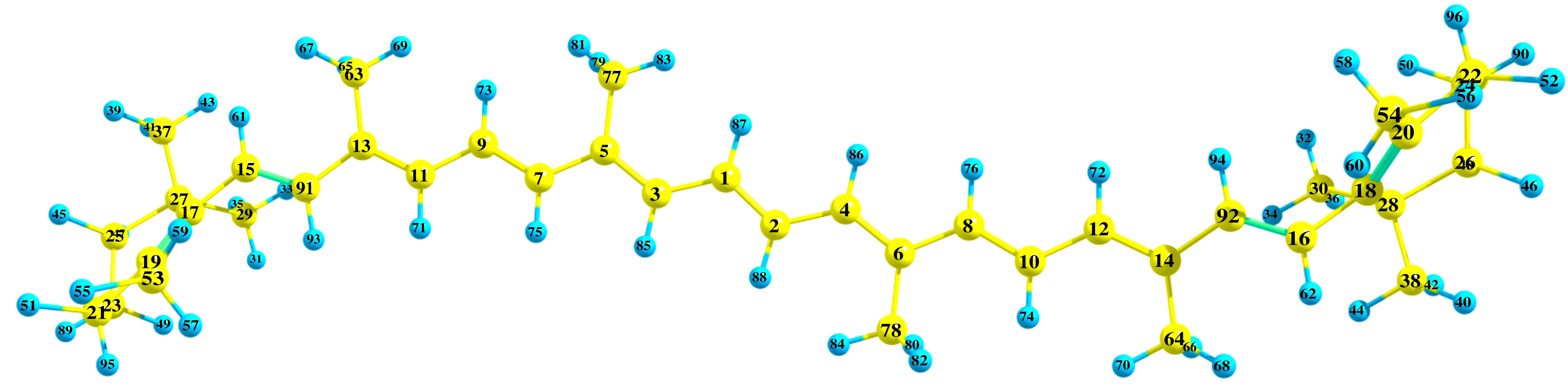}\caption{The structure and atom numeration of the C2v symmetry-carotene. \label{fig:bcar-structure}}
\end{figure*}

Temperature of the normal vibrational modes, as well as, the phonon
modes, is included by performing the Monte Carlo simulation to generate
the thermal ensemble of the Davydov $\text{D}_{2}$ wavefunction trajectories.
At the zero time, before optical excitation, in each trajectory, the
initial coherent state displacements $\tilde{\lambda}_{q}\left(0\right)$,
$\lambda_{p}\left(0\right)$ are sampled from the Glauber-Sudarshan
distribution \citep{Glauber1963}
\begin{equation}
\mathcal{P}\left(\lambda\right)=\mathcal{Z}^{-1}\exp\left(-\left|\lambda\right|^{2}\left(\text{e}^{\frac{\omega}{k_{\text{B}}T}}-1\right)\right),
\end{equation}
where $\mathcal{Z}$ is the partition function of a single coherent
state $|\lambda\rangle$ with the corresponding frequency $\omega$,
$k_{\text{B}}$ is the Boltzmann constant and $T$ is the model temperature.
Observables averaged over the thermal ensemble will be denotes as
$\left\langle \ldots\right\rangle $. We found $500$ trajectories
to be sufficient to obtain converged ensemble for the model of $\beta$-Car
as described in the next section.

\section{Simulation results}

\subsection{Normal modes of $\beta$-Car in $S_{0}$ and $S_{2}$ electronic
states}

We consider a model of $\beta$-Car in thermal equilibrium with a
solvent at $300\ \text{K}$. For the photon absorption process, $\beta$-Car
is described by the electronic ground state $|S_{0}\rangle\equiv|g\rangle$
and the excited state $|S_{2}\rangle\equiv|e\rangle$. The optically
dark excited state $|\text{S}_{1}\rangle$ does not directly participate
in electronic absorption process and is excluded.

The electronic Schrödinger equation of the $\beta$-Car molecule was
solved using the Density functional theory (DFT) method for the ground
electronic state $S_{0}$, and the time-dependent density functional
theory (TD-DFT) method for the electronic excited state $S_{2}$,
from which atom equilibrium positions $R_{0}^{g}$, $R_{0}^{e}$ are
acquired. The GAMESS \citep{Schmidt1993} and Gaussian-16 codes \citep{g16}
were used.

The calculation methods were based on the experience from previous
calculations of resonance RAMAN spectra of carotenoids, investigation
of dependence between the position of the $S_{0}\text{\ensuremath{\rightarrow}}S_{2}$
transition and the frequency of \textgreek{n}1 Raman band \citep{Macernis2014,Macernis2015}.
The most RAMAN intense band, \textgreek{n}1, located at around 1500
$\text{cm}^{-1}$, arises from the stretching of the C=C bonds. Previous
calculations of the \textgreek{n}1 Raman bands in the ground electronic
state $S_{0}$ were performed for another carotenoid, lycopene, using
the DFT method with B3LYP/6-31G, B3LYP/TZVP, B3LYP/6-31G(2df,p), BP86/6-31G(d),
BPW91/6-31G(d), B3P86/6-31G(d), B3PW91/6-31G(d), and SVWN/6-31G(d)
potentials \citep{Wei-Long2010}. It was shown that all methods based
on the DFT are able to perform calculation of vibrational frequencies
with an overall root-mean-square error of 34\textminus 48 $\text{cm}^{-1}$
\citep{Wong1996}. Also, it was shown that the dependence of the Raman
peak frequency shift, compared between the computed in vacuum and
experimental, are linear over the whole spectra \citep{Macernis2014},
using the B3LYP/6-311G(d,p) method, a scaling factor 0.9613 has to
be used \citep{Wong1996,Macernis2014}. On the other hand, energy
of the $\beta$-Car corresponding to the first optically allowed transition
in the gas phase was reported to be between 2.85 and 2.93 eV \citep{Macernis2015}.
This value is 0.62 eV higher than the excitation energy calculated
using TD-DFT at the B3LYP/6-311G(d,p) level (2.224 eV) \citep{Macernis2014}.
Other methods give similar result: Tamm-Dancoff approximation (TDA)
blyp/6-31G(d) -- 2.15 eV, TD b3lyp/cc-pvdz -- 2.19 eV , TD b3lyp/cc-pvtz
-- 2.21 eV.

\begin{figure}
\centering{}\includegraphics[width=0.5\textwidth]{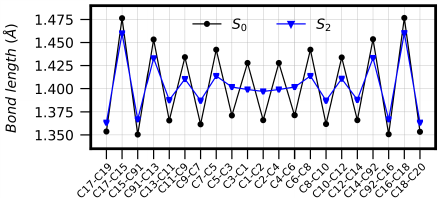}\caption{The polyene chain C-C bond lengths in the electronic ground state
$S_{0}$ and the excited state $S_{2}$ calculated using the TD-SCF
B3LYP/6-311G(d,p) method. \label{fig:polyene-chain}}
\end{figure}

Equilibrium structures of excited states, optimized using the TDA
\citep{Hirata1999} and TD-DFT \citep{Casida1995,Dreuw2005} with
the BLYP functional and DZP basis set, in contrast to B3LYP, yields
correct energetic order of the two lowest $\beta$-Car excited states,
and it has been shown to reach an accuracy of 0.2 eV for the $S_{1}$
excitation energy in carotenoids \citep{Dreuw2006}. But later this
has been explained by a fortuitous cancellation of errors caused by
the neglect of double excitations in the ground and excited states
\citep{Starcke2006}.

We performed geometry optimization using various methods, with TD-SCF,TDA-SCF
basis sets and b3lyp/6-311G(d,p), blyp/6-31G(d), b3lyp/cc-pvdz, b3lyp/cc-pvtz
potentials. The Car molecule equilibrium structure and enumeration
of atoms is shown in Fig. (\ref{fig:bcar-structure}), and the changes
of the C-C bond lengths along the Car polyene chain in both electronic
states calculated using TD-SCF B3LYP/6-311G(d,p) method is shown in
Fig. (\ref{fig:polyene-chain}). All tested methods give similar alternation
of the C-C bond lengths in $S_{0}$ state, close to that shown in
Fig. (\ref{fig:polyene-chain}). For the $S_{2}$ state, situation
is different -- methods using TDA-SCF basis set give alternation
similar to the $S_{0}$ state case. The largest alternation of the
$S_{2}$ state polyene bond lengths was achieved using the TDA blyp/6-31G(d)
method. All TD-SCF calculations provide almost 10 times smaller alternation
of C-C bond lengths in the middle of the polyene chain, as compared
to the TDA-DTF calculations, again, similar to results shown in Fig.
(\ref{fig:polyene-chain}).

\begin{figure}
\centering{}\includegraphics[width=0.48\textwidth]{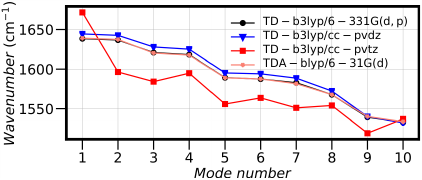}\caption{Polyene backbone C-C valence vibrational mode frequencies in the excited
state $S_{2}$ calculated using various quantum chemistry methods.
\label{fig:poliene-chain-freq}}
\end{figure}

In order to evaluate influence of the chosen method to the vibrational
mode frequencies and their bands, we performed calculation of vibrational
spectra using TD-SCF, TDA-SCF methods with different basis sets and
potentials (b3lyp/6-311G(d,p), blyp/6-31G(d), b3lyp/cc-pvdz, b3lyp/cc-pvtz).
In the $S_{0}$ state, valence vibrational frequencies of the C-C
bonds of the polyene chain scale equally and agree to within the range
of 20 $\text{cm}^{-1}$. The same vibrational frequencies in the $S_{2}$
state also scale equally, with exception of the TDA blyp/6-31G(d)
method, as shown in Fig. (\ref{fig:poliene-chain-freq}). All tested
methods agree on the C-C bond vibrational frequencies in the $S_{2}$
state within a range of 46 $\text{cm}^{-1}$. Previous \textgreek{n}1
Raman band evaluation and correlation with $S_{0}\text{\ensuremath{\rightarrow}}S_{2}$
excitation in gas phase were performed in vacuum using TD-SCF B3LYP/6-311G(d,p)
method and the results shown agreement with the experimental observations
\citep{Macernis2014,Macernis2015}. Based on this knowledge, all further
presented quantum chemistry calculations were performed in vacuum
using B3LYP/6-311G(d,p) method as in Ref. \citep{Macernis2015} and
scaling factor was not applied.

\begin{comment}
basis set/exchange correlation potential 6-311G(d,p)/b3lyp was used
to calculate potential energy force fields. Geometry of the molecule
was optimized at the same calculations level using C1 and C2 symmetry
for both electronic states. Solving eigenvalue equation for the Hessian
matrix in electronic ground and excited states, vibrational normal
modes of both electronic states were obtained. Then the excited state
normal modes were expressed using the ground state normal modes as
described in previous section. The $\beta$-Car molecule equilibrium
structure and enumeration of atoms is shown in Fig. (\ref{fig:bcar-structure}),
and the changes of the C-C bond lengths along the $\beta$-Car polyene
chain in both electronic states is shown in Fig (\ref{fig:polyene-chain}).
During geometry relaxation in $S_{2}$ state, bond lengths of the
polyene chain changes. These deformations mainly affect the $\pi$
electron system and induce changes to the normal modes during the
$S_{0}\text{\ensuremath{\rightarrow}}S_{2}$ excitation.
\end{comment}

The changes in polyene chain geometry during $S_{0}\text{\ensuremath{\rightarrow}}S_{2}$
electronic excitation causes changes in molecular electronic structure,
normal mode frequencies and vibrational mode coordinates. The C--H
valence bond vibrations in all $\beta$-Car parts are in the region
of 2970--3170 $\text{cm}^{-1}$ for the ground state, and in the
region of 2960-3168 $\text{cm}^{-1}$ for the excited state. The lower
frequency region is characterized by the change of C=C bond lengths
in polyene chain. Here, the $S_{0}$ normal mode frequencies lay in
the region of 1558-1674 $\text{cm}^{-1}$ and the corresponding region
for $S_{2}$ state is 1533-1636 $\text{cm}^{-1}$. Transition $S_{0}\rightarrow S_{2}$
mainly induces differences in the polyene chain bond lengths between
carbon atoms in both electronic states. As a consequence, vibrational
frequencies in the $S_{2}$ state become lower by 40-50 $\text{cm}^{-1}$.
\begin{figure}
\begin{centering}
\includegraphics[width=0.48\textwidth]{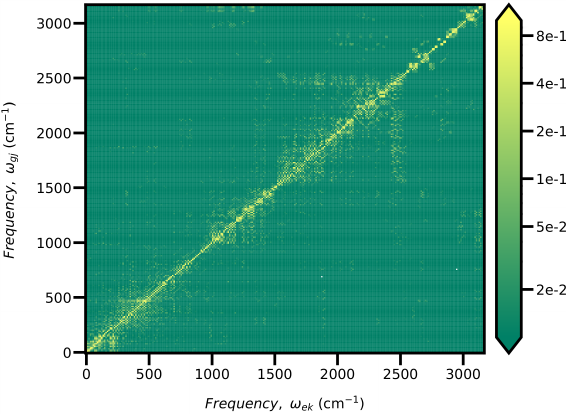}
\par\end{centering}
\caption{\label{fig:exp-coeff} Expansion coefficient absolute value $\left|a_{kj}^{(eg)}\right|$
of the $\text{\ensuremath{\beta}}$-Car normal modes. The $k$th mode
in the $S_{2}$ state in expanded in terms of the mode $j$th in the
state $S_{0}$ calculated using the TD-SCF B3LYP/6-311G(d,p) method.}
\end{figure}

Expressing normal mode coordinates in the electronic state $S_{2}$
by normal coordinates of the ground electronic state $S_{0}$, according
to the Eq. (\ref{eq:nm-expr}), allows to investigate normal mode
mixing upon $S_{0}\rightarrow S_{2}$ electronic transition. In Fig.
(\ref{fig:exp-coeff}) we plot expansion coefficient absolute value
$\left|a_{kj}^{(eg)}\right|$, i.e., the $k$th mode in the $S_{2}$
state in expanded in terms of the mode $j$th in the state $S_{0}$.
Largest expansion coefficients lay close the main diagonal, implying
that the majority of normal modes are non-negligibly mixed with similar
frequency modes. However, certain modes show mixing with modes that
has a vastly different frequencies, e.g., modes in a frequency region
of $\approx2500\ \text{cm}^{-1}$ are highly mixed with modes in a
frequency range of $1500-2500\ \text{cm}^{-1}$. Strong mixing can
also be clearly seen between modes in frequency regions of $0-250\ \text{cm}^{-1}$,
$400-750\ \text{cm}^{-1}$, $1000-1500\ \text{cm}^{-1}$. Such a broad
frequency mixing range signifies wide range of available vibrational
relaxation pathways. At a first glance, expansion coefficients along
the $\omega_{ei}=\omega_{gi}$ diagonal may look symmetric, however,
they are not, even when absolute values are considered $\left|a_{kj}^{(eg)}\right|\neq\left|a_{jk}^{(eg)}\right|$.
This demonstrates that there is no one-to-one correspondence between
the $\beta$-Car normal modes in electronic $S_{0}$ and $S_{2}$
states.

Additionaly, we found that during transitions between $S_{0}$ and
$S_{2}$ electronic states, transition dipole moment remain comparable.
For transition $S_{0}\text{\ensuremath{\rightarrow}}S_{2}$, the transition
moment components are $\mu_{02}=\left(8.35,0.71,0.0\right)$ in a.u.
$\left(21.28\ \text{Debye}\right)$, while for the $S_{2}\text{\ensuremath{\rightarrow}}S_{0}$,
it is equal to $\mu_{20}=\left(9.45,0.57,0.0\right)$ in a.u. $\left(24.06\ \text{Debye}\right)$.
Difference between the two transitions are minimal, thus non-Condon
effects can be reasonably excluded from calculations. Transition dipole
moment is oriented with z component being perpendicular to the plane
of polyene chain, while x component is directed along the polyene
chain.

\subsection{Absorption spectrum of the $\text{\ensuremath{\beta}-carotene}$
model}

The quantum chemistry results of the $\text{\ensuremath{\beta}}$-Car
is now used to compute the absorption spectrum given by the Eq. (\ref{eq:abs-fun}).
The Fourier transformation is performed on the linear response function
averaged over the thermal ensemble, $\left\langle F\left(t\right)\right\rangle $,
single trajectory of the ensemble linear response is defined in the
Eq. (\ref{eq:lin-resp-fun}), and equal to
\begin{multline}
F\left(t\right)=\left|\bm{\mu}_{eg}^{(\text{el})}\right|^{2}\text{e}^{\frac{\text{i}}{2}\sum_{k}\omega_{gk}t}\alpha_{g}^{\star}\left(0\right)\alpha_{e}\left(t\right)\\
\times\text{\ensuremath{\exp}\ensuremath{\sum_{k}}\ensuremath{\left(\text{e}^{\text{i}\omega_{gk}t}\tilde{\lambda}_{k}^{\star}\left(0\right)\tilde{\lambda}_{k}\left(t\right)-\frac{1}{2}\left(\left|\tilde{\lambda}_{k}\left(0\right)\right|^{2}+\left|\tilde{\lambda}_{k}\left(t\right)\right|^{2}\right)\right)}}\\
\times\text{\ensuremath{\exp}\ensuremath{\sum_{p}}\ensuremath{\left(\text{e}^{\text{i}w_{p}t}\lambda_{p}^{\star}\left(0\right)\lambda_{p}\left(t\right)-\frac{1}{2}\left(\left|\lambda_{p}\left(0\right)\right|^{2}+\left|\lambda_{p}\left(t\right)\right|^{2}\right)\right)}}.\label{eq:LRF}
\end{multline}
It is expressed in terms of the dynamical parameters $\alpha_{n}$$\left(t\right)$,
$\tilde{\lambda}_{k}\left(t\right)$, $\lambda_{p}\left(t\right)$,
therefore it is enough to propagate the excited state dynamics.

For the solvent, the phonon bath modes are defined by uniformly discretizing
the spectral density function $C_{e}^{"}\left(\omega\right)$ in frequency
domain in the range $\left[w_{\text{min}}=0.1,\ w_{\text{max}}=1250\right]\ \text{cm}^{-1}$
with discretization step size $\Delta_{w}=10\ \text{cm}^{-1}$. Then
the frequency of $p$th bath mode is given by $w_{p}=w_{\text{min}}+\left(p-1\right)\Delta_{w}$.
Form of the spectral density function was chosen to be the Overdamped
Brownian oscillator function $C_{e}^{"}\left(w\right)=2\Lambda_{e}^{\mathrm{ph}}\omega\gamma/(\omega^{2}+\gamma^{2})$.
Damping parameter $\gamma=200\ \text{cm}^{-1}$ $\left(167\ \text{fs}\right)$
has been chosen based on the previous modeling of $\beta$-Car \citep{Balevicius2015}
spectra. Amplitude of the spectral density function is set by normalizing
$f_{ep}$ values according to the reorganization energy definition
by the Eq. (\ref{eq:Lphon}). The bath reorganization energy of $\Lambda_{e}^{\text{ph}}=100\ \text{cm}^{-1}$
have been chosen to qualitatively match the line widths of the experimental
data.

\begin{figure}
\begin{centering}
\includegraphics[width=0.48\textwidth]{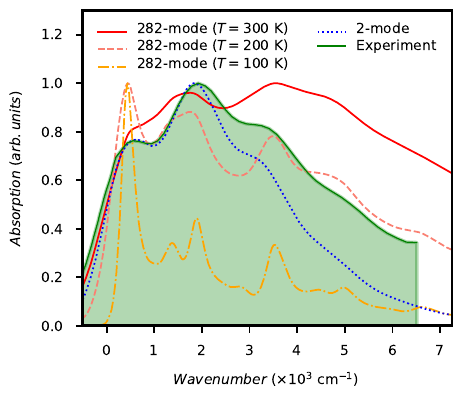}
\par\end{centering}
\caption{Absorption spectra of $\text{\ensuremath{\beta}}$-Car model, based
on B3LYP/6-311G(d,p) method, at different temperatures along the experimental
$\text{\ensuremath{\beta}}$-Car spectrum in diethylamine solvent
at room temperature (shown as contoured green). The widely used 2-mode
model at 300 $\text{K}$ temperature is also shown for comparison,
model parameters are taken from Ref. \citep{Polivka2001}. All spectra
are normalized to their maximum value and aligned on their 0-0 transition
band.\label{fig:Abs-eff}}
\end{figure}

The simulated absorption spectrum of $\text{\ensuremath{\beta}}$-Car
model with 282 normal modes at different temperatures is shown in
Fig. (\ref{fig:Abs-eff}) along the experimental $\text{\ensuremath{\beta}}$-Car
spectrum in diethylamine solvent at room temperature \citep{Gong2018}.
Absorption spectra have been normalized to their maximum value, as
well as, aligned on the 0-0 transition band for easier comparison.
We find the 282-mode model spectrum to qualitatively reproduce position
and amplitudes of the first two absorption peaks, however, it greatly
overestimates the amplitude of vibrational peak progression at $300$
$\text{K}$ temperature. Also, absorption of the high frequency modes
display non-trivial dependence on the temperature. For majority of
modes the average thermal energy is much smaller than the energy gap
between the vibrational mode energy levels, $k_{\text{B}}T\ll\omega$,
thus, for non-mixed modes, dependence of absorption spectrum on temperature
would be negligible. However, in Fig. (\ref{fig:Abs-eff}) we observe
strong dependence of absorption on temperature due to the mode mixing,
i.e., thermally excited low frequency vibrational modes contribute
to the excitation of the high frequency modes, which results in a
wide high frequency absorption shoulder.

For comparison, we also computed absorption spectrum of a widely used
empirical 2-mode $\text{\ensuremath{\beta}}$-Car model at $300$
$\text{K}$ temperature, which includes only the $\text{C=C}$ and
$\text{C-C}$ stretching vibrational modes with no mixing between
them. Typical model frequencies $\omega_{\text{e,C=C}}=1522\ \text{cm}^{-1}$,
$\omega_{e,\text{C-C}}=1157\ \text{cm}^{-1}$ and displacements $d_{\text{C=C}}^{\left(eg\right)}=1.3$,
$d_{\text{C-C}}^{\left(eg\right)}=0.9$ are taken from Ref. \citep{Polivka2001}.
To have correct line widths, bath reorganization energy is now set
to a much larger $\Lambda_{e}^{\text{ph}}=800\ \text{cm}^{-1}$, this
is to account for the lack of the rest $\beta$-Car modes. As shown
in Fig. (\ref{fig:Abs-eff}), the 2-mode model fits first two peaks
well, but underestimates amplitude of the higher frequency progression.

\begin{figure}
\begin{centering}
\includegraphics[width=0.48\textwidth]{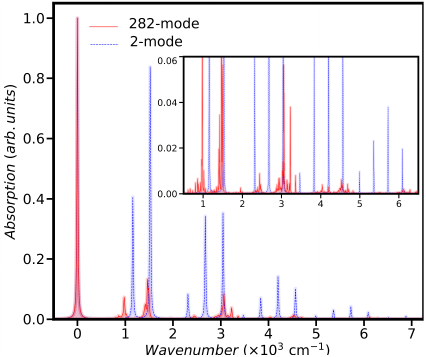}
\par\end{centering}
\caption{Stick absorption spectrum of the 282-mode model, computed using the
B3LYP/6-311G(d,p) method, and the empirical 2-mode model. Purely electronic
transition energy is set to $0\ \text{cm}^{-1}$ for both spectra.
For visibility, each spectra have been convoluted with the $\tau=1\ \text{ps}$
variance Gaussian function. Inset more closely shows low amplitude
sticks, and these has been convoluted with $\tau=5\ \text{ps}$ variance
Gaussian function. \label{fig:Abs-stick-zero}}
\end{figure}

To further compare the 282-mode and the 2-mode models, we look at
their stick absorption spectrum in Fig. (\ref{fig:Abs-stick-zero}).
The purely electronic transition energy is set to $0\ \text{cm}^{-1}$
for both spectra. For visibility, spectra have been convoluted with
the $\tau=1\ \text{ps}$ variance Gaussian function, $\tau=5\ \text{ps}$
is used for spectra in the inset. The 2-mode model stick spectrum
has a straightforward peak progression, i.e., spectrum is a sum of
each of the two mode peak progressions. The 282-mode model spectrum
has a more complex structure. Even though each of the 282 modes have
a small absorption peak, the combined spectrum produces frequency
regions with non-negligible absorption intensity. These regions show
clear overlap with the absorption peaks of the 2-mode model. The 2-mode
spectrum has peaks at $1522\ \text{cm}^{-1}$ and $1157\ \text{cm}^{-1}$
frequencies, produced by the $\text{C=C}$ and $\text{C-C}$ stretching
vibrational modes. The 282-mode spectrum has similar frequency regions,
only this time, they are due to the absorption of a large number of
mixed normal modes. These modes are responsible for the first two
peaks seen in Fig. (\ref{fig:Abs-eff}) spectra.

Looking further on in Fig. (\ref{fig:Abs-stick-zero}), the 282-mode
model has absorption in the $3000\ \text{cm}^{-1}$, $4500\ \text{cm}^{-1}$,
$6000\ \text{cm}^{-1}$ frequency regions. These account for the high
frequency absorption tail seen in experiments. Due to the $\text{C=C}$
and $\text{C-C}$ mode progression, the 2-mode model has a peak at
these frequencies as well, however, even though visually they look
more intese than the 282-mode model peaks, Fig. (\ref{fig:Abs-eff})
simulations show it being the opposite. Again, the strong absorption
is produced by the summation of a large number of weak intensity absorption
peaks. Two harmonic modes simply can not accurately describe absorption
over a such a wide range of frequencies, therefore, the high frequency
absorption of the 2-mode model is lacking.

\section{Discussion}

Vibrational modes of carotenoids have been extensively studied by
Raman spectroscopy \citep{DeOliveira2009,Tschirner2009a}. The frequency
of the most Raman active V1 vibration in the $S_{0}$ state is 1642.3
$\text{cm}^{-1}$, which changes to a week Raman vibration of frequency
1584.08 $\text{cm}^{-1}$ in the $S_{2}$ state. The C-C valence bond
frequencies in $S_{0}$ state lay in the region of 1018-1353 $\text{cm}^{-1}$,
while it is in region of 1156-1370 $\text{cm}^{-1}$ for the $S_{2}$
state. These frequencies are strongly mixed with the polyene chain
C-H bond in-plane vibrations, and the C-C valence vibrations of peripheral
rings. The strongest V2 Raman active vibration in this region, for
the $S_{0}$ state, is 1187.22 $\text{cm}^{-1}$, while the mode with
the most similar vibrational form in the $S_{2}$ state has a frequency
of 1219 $\text{cm}^{-1}$. Vibrations of lower frequencies are associated
with the C-H vibrations outside of the polyene chain, deformations
of the peripheral rings, changes of the polyene chain valence angles,
dihedral angles and the deformations of the whole molecule by twisting
and waving. Frequencies of these vibrations change by no more than
10 $\text{cm}^{-1}$ after the $S_{0}\rightarrow S_{2}$ transition.
Difference in the vibrational forms are not as strong for these modes,
as were in the case of the polyene chain C-C and C=C valence bond
vibrations.

Recently Balevi\v{c}ius Jr. and co-workers \citep{BaleviciusJr2019a}
have presented an in-depth excitation energy relaxation model in carotenoids
by considering four relaxation processes. Simply put, the event of
photoexcitation instantaneously promotes the carotenoid molecules
to a non-equilibrium state and launches the internal vibrational redistribution
(IVR) cascade within the high-frequency optically active modes resulting
into transient thermally \textquotedbl hot\textquotedbl{} state.
Generally, it is assumed that the thermally hot carotenoid subsequently
transfers vibrational energy to the solvent molecules, i.e., vibrational
cooling (VC) takes place. Authors demonstrated how modeling the IVR
and VC concurrently, and not subsequently, naturally explains the
presence of the highly discussed transient absorption $S^{*}$ signal
\citep{Polivka2009} in terms of the vibrationally hot ground state
$S_{0}$. The 2-mode model was used, i.e., only the $\text{C-C}$
and $\text{C=C}$ intramolecular modes were coupled to the thermal
bath - their coupling strength remains speculative. Both the IVC and
VC relaxation were modeled implicitly by prescribing process timescales.
We have shown that the 2-mode (not mixed) model is not sufficient
in describing the photon absorption spectrum. In fact, upon photoexcitation
many vibrational modes become excited. No two distinctive modes could
be isolated in the relevant frequency region. We observe grouping
of vibrational modes in the $1000\ \text{cm}^{-1}$ and $1500\ \text{cm}^{-1}$
frequency regions, as shown in Fig. \ref{fig:Abs-stick-zero}. The
2-mode model yields progression of peaks with few strong features
at $2700\ \text{cm}^{-1}$ and $4000-6000\ \text{cm}^{-1}$ frequencies.
Meanwhile, the 282-mode model has a large number of weak absorption
features at these frequencies, however, it is the cumulative effect
that combines them into the observed \textquotedbl progression\textquotedbl .
This entropic factor actually simplifies the overall electronic excitation
relaxation picture since each mode is weakly coupled to the electronic
transition, therefore, the weak coupling regime could be used in theoretical
models of relaxation dynamics. Consequently, the two- or multiple-quanta
vibrational excitations become improbable. Hence, only the entropic
factor as a cumulative effect of all modes would have decisive impact
on both the IVR and VC processes timescales.

In Nature carotenoids participate in energy conversion process together
with other types of pigments. Carotenoids play an important role in
light-harvesting complexes by transfers their excitation to chlorophylls
on a femtosecond timescale. It is especially evident in the peridinin--chlorophyll
a protein (PCP), in which the dominant energy transfer occurs from
the peridinin $S_{2}$ to chlorophyll $Q_{y}$ state via an ultrafast
coherent mechanism. The coherent superposition of the two states functions
in a way as to drive the population to the final acceptor state \citep{Roscioli2017},
providing an important piece of evidence in the quest of connecting
coherent phenomena and biological functions \citep{Meneghin2018a}.
This process is highly sensitive to structural perturbations of the
peridinin polyene backbone, which has a profound effect on the overall
lifetime of the complex \citep{Ghosh2017}. We have found that $\beta$-Car
as well undergoes polyene backbone changes, mainly, in its C-C bond
lengths.

As well, it has been suggested that the ultrafast population transfer
from the carotenoid state $S_{2}$ to the bacteriochlorophyll (BChl)
state $Q_{x}$ occurs due to the vibronic coupling of the carotenoid
electron-vibrational degrees of freedom to the BChl \citep{Perlik2015b}.
Energy flow pathway opened up by the resonance of the energy gap between
the carotenoid vibrational levels, and the BChl $|g\rangle_{\text{BChl}}\rightarrow|Q_{x}\rangle$
transition is the primary reason for its ultrafast nature. We, hence,
suggest, that by going beyond the 2-mode model and taking into account
more carotenoid vibrational modes, in turn, more vibrational levels,
the probability of resonance between the carotenoid and BChl would
greatly increased, changing the overall population transfer rate.

In conclusion, we have presented a $\beta$-Car model with fully explicit
treatment of all its 282 vibrational normal modes, which were computed
using the quantum chemical methods. Additionally, we described how
to treat the $\beta$-Car excited states dynamics when in contact
with solvent at finite temperature. We found $\beta$-Car to change
bond lengths between the polyene chain atoms during the $S_{0}\rightarrow S_{2}$
electronic transition, as well as, that the there is no one-to-one
correspondence between the ground and the excited state vibrational
modes, i.e., modes on different electronic states are highly mixed
and should not be treated as being the same. Model absorption spectrum
qualitatively match the experimental data, it better describe the
high frequency progression of the carotenoid spectrum than the typical
2-mode carotenoid model.
\begin{suppinfo}
Derivation of the Davydov $D_{2}$ ansatz equations of motion for
the $\beta$-carotene model using Dirac-Frenkel variational method.
\end{suppinfo}
\begin{acknowledgement}
We thank the Research Council of Lithuania for financial support (grant
No: S-MIP-20-47). Computations were performed on resources at the
High Performance Computing Center, ``HPC Sauletekis'' in Vilnius
University Faculty of Physics.
\end{acknowledgement}
\bibliographystyle{achemso}
\bibliography{JMD_with_pages}

\providecommand{\latin}[1]{#1}
\makeatletter
\providecommand{\doi}
  {\begingroup\let\do\@makeother\dospecials
  \catcode`\{=1 \catcode`\}=2 \doi@aux}
\providecommand{\doi@aux}[1]{\endgroup\texttt{#1}}
\makeatother
\providecommand*\mcitethebibliography{\thebibliography}
\csname @ifundefined\endcsname{endmcitethebibliography}
  {\let\endmcitethebibliography\endthebibliography}{}
\begin{mcitethebibliography}{75}
\providecommand*\natexlab[1]{#1}
\providecommand*\mciteSetBstSublistMode[1]{}
\providecommand*\mciteSetBstMaxWidthForm[2]{}
\providecommand*\mciteBstWouldAddEndPuncttrue
  {\def\EndOfBibitem{\unskip.}}
\providecommand*\mciteBstWouldAddEndPunctfalse
  {\let\EndOfBibitem\relax}
\providecommand*\mciteSetBstMidEndSepPunct[3]{}
\providecommand*\mciteSetBstSublistLabelBeginEnd[3]{}
\providecommand*\EndOfBibitem{}
\mciteSetBstSublistMode{f}
\mciteSetBstMaxWidthForm{subitem}{(\alph{mcitesubitemcount})}
\mciteSetBstSublistLabelBeginEnd
  {\mcitemaxwidthsubitemform\space}
  {\relax}
  {\relax}

\bibitem[Tian(2015)]{Tian2015}
Tian,~L. {Recent advances in understanding carotenoid-derived signaling
  molecules in regulating plant growth and development}. \emph{Front. Plant
  Sci.} \textbf{2015}, \emph{6}, 790\relax
\mciteBstWouldAddEndPuncttrue
\mciteSetBstMidEndSepPunct{\mcitedefaultmidpunct}
{\mcitedefaultendpunct}{\mcitedefaultseppunct}\relax
\EndOfBibitem
\bibitem[Weaver \latin{et~al.}(2018)Weaver, Santos, Tucker, Wilson, and
  Hill]{Weaver2018}
Weaver,~R.~J.; Santos,~E.~S.; Tucker,~A.~M.; Wilson,~A.~E.; Hill,~G.~E.
  {Carotenoid metabolism strengthens the link between feather coloration and
  individual quality}. \emph{Nat. Commun.} \textbf{2018}, \emph{9}, 73\relax
\mciteBstWouldAddEndPuncttrue
\mciteSetBstMidEndSepPunct{\mcitedefaultmidpunct}
{\mcitedefaultendpunct}{\mcitedefaultseppunct}\relax
\EndOfBibitem
\bibitem[Murchie and Harbinson(2014)Murchie, and Harbinson]{Murchie2014}
Murchie,~E.~H.; Harbinson,~J. \emph{{Non-Photochemical Fluorescence Quenching
  Across Scales: From Chloroplasts to Plants to Communities}}; Springer,
  Dordrecht, 2014; pp 553--582\relax
\mciteBstWouldAddEndPuncttrue
\mciteSetBstMidEndSepPunct{\mcitedefaultmidpunct}
{\mcitedefaultendpunct}{\mcitedefaultseppunct}\relax
\EndOfBibitem
\bibitem[Ruban(2016)]{Ruban2016}
Ruban,~A.~V. {Nonphotochemical chlorophyll fluorescence quenching: Mechanism
  and effectiveness in protecting plants from photodamage}. \emph{Plant
  Physiol.} \textbf{2016}, \emph{170}, 1903--1916\relax
\mciteBstWouldAddEndPuncttrue
\mciteSetBstMidEndSepPunct{\mcitedefaultmidpunct}
{\mcitedefaultendpunct}{\mcitedefaultseppunct}\relax
\EndOfBibitem
\bibitem[Ruban \latin{et~al.}(2007)Ruban, Berera, Ilioaia, {Van Stokkum},
  Kennis, Pascal, {Van Amerongen}, Robert, Horton, and {Van
  Grondelle}]{Ruban2007}
Ruban,~A.~V.; Berera,~R.; Ilioaia,~C.; {Van Stokkum},~I.~H.; Kennis,~J.~T.;
  Pascal,~A.~A.; {Van Amerongen},~H.; Robert,~B.; Horton,~P.; {Van
  Grondelle},~R. {Identification of a mechanism of photoprotective energy
  dissipation in higher plants}. \emph{Nature} \textbf{2007}, \emph{450},
  575--578\relax
\mciteBstWouldAddEndPuncttrue
\mciteSetBstMidEndSepPunct{\mcitedefaultmidpunct}
{\mcitedefaultendpunct}{\mcitedefaultseppunct}\relax
\EndOfBibitem
\bibitem[Holt \latin{et~al.}(2005)Holt, Zigmantas, Valkunas, Li, Niyogi, and
  Fleming]{Holt2005}
Holt,~N.~E.; Zigmantas,~D.; Valkunas,~L.; Li,~X.~P.; Niyogi,~K.~K.;
  Fleming,~G.~R. {Carotenoid cation formation and the regulation of
  photosynthetic light harvesting}. \emph{Science} \textbf{2005}, \emph{307},
  433--436\relax
\mciteBstWouldAddEndPuncttrue
\mciteSetBstMidEndSepPunct{\mcitedefaultmidpunct}
{\mcitedefaultendpunct}{\mcitedefaultseppunct}\relax
\EndOfBibitem
\bibitem[Llansola-Portoles \latin{et~al.}(2017)Llansola-Portoles, Sobotka,
  Kish, Shukla, Pascal, Pol{\'{i}}vka, and Robert]{Llansola-Portoles2017}
Llansola-Portoles,~M.~J.; Sobotka,~R.; Kish,~E.; Shukla,~M.~K.; Pascal,~A.~A.;
  Pol{\'{i}}vka,~T.; Robert,~B. {Twisting a $\beta$-carotene, an adaptive trick
  from nature for dissipating energy during photoprotection}. \emph{J. Biol.
  Chem.} \textbf{2017}, \emph{292}, 1396--1403\relax
\mciteBstWouldAddEndPuncttrue
\mciteSetBstMidEndSepPunct{\mcitedefaultmidpunct}
{\mcitedefaultendpunct}{\mcitedefaultseppunct}\relax
\EndOfBibitem
\bibitem[Bode \latin{et~al.}(2009)Bode, Quentmeier, Liao, Hafi, Barros, Wilk,
  Bittner, and Walla]{Bode2009}
Bode,~S.; Quentmeier,~C.~C.; Liao,~P.~N.; Hafi,~N.; Barros,~T.; Wilk,~L.;
  Bittner,~F.; Walla,~P.~J. {On the regulation of photosynthesis by excitonic
  interactions between carotenoids and chlorophylls}. \emph{Proc. Natl. Acad.
  Sci. U. S. A.} \textbf{2009}, \emph{106}, 12311--12316\relax
\mciteBstWouldAddEndPuncttrue
\mciteSetBstMidEndSepPunct{\mcitedefaultmidpunct}
{\mcitedefaultendpunct}{\mcitedefaultseppunct}\relax
\EndOfBibitem
\bibitem[Cupellini \latin{et~al.}(2020)Cupellini, Calvani, Jacquemin, and
  Mennucci]{Cupellini2020}
Cupellini,~L.; Calvani,~D.; Jacquemin,~D.; Mennucci,~B. {Charge transfer from
  the carotenoid can quench chlorophyll excitation in antenna complexes of
  plants}. \emph{Nat. Commun.} \textbf{2020}, \emph{11}, 1--8\relax
\mciteBstWouldAddEndPuncttrue
\mciteSetBstMidEndSepPunct{\mcitedefaultmidpunct}
{\mcitedefaultendpunct}{\mcitedefaultseppunct}\relax
\EndOfBibitem
\bibitem[Pol{\'{i}}vka and Sundstr{\"{o}}m(2004)Pol{\'{i}}vka, and
  Sundstr{\"{o}}m]{Polivka2004}
Pol{\'{i}}vka,~T.; Sundstr{\"{o}}m,~V. {Ultrafast dynamics of carotenoid
  excited states-from solution to natural and artificial systems}. \emph{Chem.
  Rev.} \textbf{2004}, \emph{104}, 2021--2071\relax
\mciteBstWouldAddEndPuncttrue
\mciteSetBstMidEndSepPunct{\mcitedefaultmidpunct}
{\mcitedefaultendpunct}{\mcitedefaultseppunct}\relax
\EndOfBibitem
\bibitem[Llansola-Portoles \latin{et~al.}(2017)Llansola-Portoles, Pascal, and
  Robert]{Llansola-Portoles2017a}
Llansola-Portoles,~M.~J.; Pascal,~A.~A.; Robert,~B. {Electronic and vibrational
  properties of carotenoids: from in vitro to in vivo}. \emph{J. R. Soc.
  Interface} \textbf{2017}, \emph{14}, 20170504\relax
\mciteBstWouldAddEndPuncttrue
\mciteSetBstMidEndSepPunct{\mcitedefaultmidpunct}
{\mcitedefaultendpunct}{\mcitedefaultseppunct}\relax
\EndOfBibitem
\bibitem[Bautista \latin{et~al.}(1999)Bautista, Connors, Raju, Hiller,
  Sharples, Gosztola, Wasielewski, and Frank]{Bautista1999}
Bautista,~J.~A.; Connors,~R.~E.; Raju,~B.~B.; Hiller,~R.~G.; Sharples,~F.~P.;
  Gosztola,~D.; Wasielewski,~M.~R.; Frank,~H.~A. {Excited State Properties of
  Peridinin: Observation of a Solvent Dependence of the Lowest Excited Singlet
  State Lifetime and Spectral Behavior Unique among Carotenoids}. \emph{J.
  Phys. Chem. B} \textbf{1999}, \emph{103}, 8751--8758\relax
\mciteBstWouldAddEndPuncttrue
\mciteSetBstMidEndSepPunct{\mcitedefaultmidpunct}
{\mcitedefaultendpunct}{\mcitedefaultseppunct}\relax
\EndOfBibitem
\bibitem[Frank \latin{et~al.}(2000)Frank, Bautista, Josue, Pendon, Hiller,
  Sharples, Gosztola, and Wasielewski]{Frank2000}
Frank,~H.~A.; Bautista,~J.~A.; Josue,~J.; Pendon,~Z.; Hiller,~R.~G.;
  Sharples,~F.~P.; Gosztola,~D.; Wasielewski,~M.~R. {Effect of the Solvent
  Environment on the Spectroscopic Properties and Dynamics of the Lowest
  Excited States of Carotenoids}. \emph{J. Phys. Chem. B} \textbf{2000},
  \emph{104}, 4569--4577\relax
\mciteBstWouldAddEndPuncttrue
\mciteSetBstMidEndSepPunct{\mcitedefaultmidpunct}
{\mcitedefaultendpunct}{\mcitedefaultseppunct}\relax
\EndOfBibitem
\bibitem[Tamm(1991)]{Tamm1945}
Tamm,~I. {Relativistic Interaction of Elementary Particles}. \emph{Sel. Pap.}
  \textbf{1991}, \emph{9}, 157--174\relax
\mciteBstWouldAddEndPuncttrue
\mciteSetBstMidEndSepPunct{\mcitedefaultmidpunct}
{\mcitedefaultendpunct}{\mcitedefaultseppunct}\relax
\EndOfBibitem
\bibitem[Dancoff(1950)]{Dancoff1950}
Dancoff,~S.~M. {Non-adiabatic meson theory of nuclear forces}. \emph{Phys.
  Rev.} \textbf{1950}, \emph{78}, 382--385\relax
\mciteBstWouldAddEndPuncttrue
\mciteSetBstMidEndSepPunct{\mcitedefaultmidpunct}
{\mcitedefaultendpunct}{\mcitedefaultseppunct}\relax
\EndOfBibitem
\bibitem[Andreussi \latin{et~al.}(2015)Andreussi, Knecht, Marian, Kongsted, and
  Mennucci]{Andreussi2015}
Andreussi,~O.; Knecht,~S.; Marian,~C.~M.; Kongsted,~J.; Mennucci,~B.
  {Carotenoids and light-harvesting: From DFT/MRCI to the Tamm-Dancoff
  approximation}. \emph{J. Chem. Theory Comput.} \textbf{2015}, \emph{11},
  655--666\relax
\mciteBstWouldAddEndPuncttrue
\mciteSetBstMidEndSepPunct{\mcitedefaultmidpunct}
{\mcitedefaultendpunct}{\mcitedefaultseppunct}\relax
\EndOfBibitem
\bibitem[Vaswani \latin{et~al.}(2003)Vaswani, Hsu, Head-Gordon, and
  Fleming]{Vaswani2003}
Vaswani,~H.~M.; Hsu,~C.~P.; Head-Gordon,~M.; Fleming,~G.~R. {Quantum chemical
  evidence for an intramolecular charge-transfer state in the carotenoid
  peridinin of peridinin-chlorophyll-protein}. \emph{J. Phys. Chem. B}
  \textbf{2003}, \emph{107}, 7940--7946\relax
\mciteBstWouldAddEndPuncttrue
\mciteSetBstMidEndSepPunct{\mcitedefaultmidpunct}
{\mcitedefaultendpunct}{\mcitedefaultseppunct}\relax
\EndOfBibitem
\bibitem[Hashimoto \latin{et~al.}(2018)Hashimoto, Uragami, Yukihira, Gardiner,
  and Cogdell]{Hashimoto2018}
Hashimoto,~H.; Uragami,~C.; Yukihira,~N.; Gardiner,~A.~T.; Cogdell,~R.~J.
  {Understanding/unravelling carotenoid excited singlet states}. \emph{J. R.
  Soc. Interface} \textbf{2018}, \emph{15}, 20180026\relax
\mciteBstWouldAddEndPuncttrue
\mciteSetBstMidEndSepPunct{\mcitedefaultmidpunct}
{\mcitedefaultendpunct}{\mcitedefaultseppunct}\relax
\EndOfBibitem
\bibitem[Premvardhan \latin{et~al.}(2005)Premvardhan, Papagiannakis, Hiller,
  and {Van Grondelle}]{Premvardhan2005}
Premvardhan,~L.; Papagiannakis,~E.; Hiller,~R.~G.; {Van Grondelle},~R. {The
  charge-transfer character of the S0 {$\rightarrow$} S2 transition in the
  carotenoid peridinin is revealed by stark spectroscopy}. \emph{J. Phys. Chem.
  B} \textbf{2005}, \emph{109}, 15589--15597\relax
\mciteBstWouldAddEndPuncttrue
\mciteSetBstMidEndSepPunct{\mcitedefaultmidpunct}
{\mcitedefaultendpunct}{\mcitedefaultseppunct}\relax
\EndOfBibitem
\bibitem[Balevi{\v{c}}ius \latin{et~al.}(2019)Balevi{\v{c}}ius, Wei, {Di
  Tommaso}, Abramavicius, Hauer, Pol{\'{i}}vka, and Duffy]{BaleviciusJr2019a}
Balevi{\v{c}}ius,~V.; Wei,~T.; {Di Tommaso},~D.; Abramavicius,~D.; Hauer,~J.;
  Pol{\'{i}}vka,~T.; Duffy,~C.~D. {The full dynamics of energy relaxation in
  large organic molecules: From photo-excitation to solvent heating}.
  \emph{Chem. Sci.} \textbf{2019}, \emph{10}, 4792--4804\relax
\mciteBstWouldAddEndPuncttrue
\mciteSetBstMidEndSepPunct{\mcitedefaultmidpunct}
{\mcitedefaultendpunct}{\mcitedefaultseppunct}\relax
\EndOfBibitem
\bibitem[Mendes-Pinto \latin{et~al.}(2013)Mendes-Pinto, Sansiaume, Hashimoto,
  Pascal, Gall, and Robert]{Mendes-Pinto2013}
Mendes-Pinto,~M.~M.; Sansiaume,~E.; Hashimoto,~H.; Pascal,~A.~A.; Gall,~A.;
  Robert,~B. {Electronic absorption and ground state structure of carotenoid
  molecules}. \emph{J. Phys. Chem. B} \textbf{2013}, \emph{117},
  11015--11021\relax
\mciteBstWouldAddEndPuncttrue
\mciteSetBstMidEndSepPunct{\mcitedefaultmidpunct}
{\mcitedefaultendpunct}{\mcitedefaultseppunct}\relax
\EndOfBibitem
\bibitem[Wei \latin{et~al.}(2019)Wei, Balevi{\v{c}}ius, Pol{\'{i}}vka, Ruban,
  and Duffy]{Wei2019}
Wei,~T.; Balevi{\v{c}}ius,~V.; Pol{\'{i}}vka,~T.; Ruban,~A.~V.; Duffy,~C.~D.
  {How carotenoid distortions may determine optical properties: Lessons from
  the orange carotenoid protein}. \emph{Phys. Chem. Chem. Phys.} \textbf{2019},
  \emph{21}, 23187--23197\relax
\mciteBstWouldAddEndPuncttrue
\mciteSetBstMidEndSepPunct{\mcitedefaultmidpunct}
{\mcitedefaultendpunct}{\mcitedefaultseppunct}\relax
\EndOfBibitem
\bibitem[Pol{\'{i}}vka \latin{et~al.}(2001)Pol{\'{i}}vka, Zigmantas, Frank,
  Bautista, Herek, Koyama, Fujii, and Sundstr{\"{o}}m]{Polivka2001}
Pol{\'{i}}vka,~T.; Zigmantas,~D.; Frank,~H.~A.; Bautista,~J.~A.; Herek,~J.~L.;
  Koyama,~Y.; Fujii,~R.; Sundstr{\"{o}}m,~V. {Near-infrared time-resolved study
  of the S1 state dynamics of the carotenoid spheroidene}. \emph{J. Phys. Chem.
  B} \textbf{2001}, \emph{105}, 1072--1080\relax
\mciteBstWouldAddEndPuncttrue
\mciteSetBstMidEndSepPunct{\mcitedefaultmidpunct}
{\mcitedefaultendpunct}{\mcitedefaultseppunct}\relax
\EndOfBibitem
\bibitem[Christensson \latin{et~al.}(2009)Christensson, Milota, Nemeth,
  Sperling, Kauffmann, Pullerits, and Hauer]{Christensson2009}
Christensson,~N.; Milota,~F.; Nemeth,~A.; Sperling,~J.; Kauffmann,~H.~F.;
  Pullerits,~T.; Hauer,~J. {Two-dimensional electronic spectroscopy of
  $\beta$-carotene}. \emph{J. Phys. Chem. B} \textbf{2009}, \emph{113},
  16409--16419\relax
\mciteBstWouldAddEndPuncttrue
\mciteSetBstMidEndSepPunct{\mcitedefaultmidpunct}
{\mcitedefaultendpunct}{\mcitedefaultseppunct}\relax
\EndOfBibitem
\bibitem[Balevi{\v{c}}ius \latin{et~al.}(2016)Balevi{\v{c}}ius, Abramavicius,
  Pol{\'{i}}vka, {Galestian Pour}, and Hauer]{Balevicius2016}
Balevi{\v{c}}ius,~V.; Abramavicius,~D.; Pol{\'{i}}vka,~T.; {Galestian
  Pour},~A.; Hauer,~J. {A Unified Picture of S* in Carotenoids}. \emph{J. Phys.
  Chem. Lett.} \textbf{2016}, \emph{7}, 3347--3352\relax
\mciteBstWouldAddEndPuncttrue
\mciteSetBstMidEndSepPunct{\mcitedefaultmidpunct}
{\mcitedefaultendpunct}{\mcitedefaultseppunct}\relax
\EndOfBibitem
\bibitem[Fox \latin{et~al.}(2017)Fox, Balevi{\v{c}}ius, Chmeliov, Valkunas,
  Ruban, and Duffy]{Fox2017}
Fox,~K.~F.; Balevi{\v{c}}ius,~V.; Chmeliov,~J.; Valkunas,~L.; Ruban,~A.~V.;
  Duffy,~C.~D. {The carotenoid pathway: What is important for excitation
  quenching in plant antenna complexes?} \emph{Phys. Chem. Chem. Phys.}
  \textbf{2017}, \emph{19}, 22957--22968\relax
\mciteBstWouldAddEndPuncttrue
\mciteSetBstMidEndSepPunct{\mcitedefaultmidpunct}
{\mcitedefaultendpunct}{\mcitedefaultseppunct}\relax
\EndOfBibitem
\bibitem[Gong \latin{et~al.}(2018)Gong, Fu, Wang, Cao, Li, Sun, and
  Men]{Gong2018}
Gong,~N.; Fu,~H.; Wang,~S.; Cao,~X.; Li,~Z.; Sun,~C.; Men,~Z.
  {All-trans-$\beta$-carotene absorption shift and electron-phonon coupling
  modulated by solvent polarizability}. \emph{J. Mol. Liq.} \textbf{2018},
  \emph{251}, 417--422\relax
\mciteBstWouldAddEndPuncttrue
\mciteSetBstMidEndSepPunct{\mcitedefaultmidpunct}
{\mcitedefaultendpunct}{\mcitedefaultseppunct}\relax
\EndOfBibitem
\bibitem[Valkunas \latin{et~al.}(2013)Valkunas, Abramavicius, and
  Mancal]{Valkunasa}
Valkunas,~L.; Abramavicius,~D.; Mancal,~T. \emph{Molecular Excitation Dynamics
  and Relaxation}; John Wiley {\&} Sons, Ltd, 2013\relax
\mciteBstWouldAddEndPuncttrue
\mciteSetBstMidEndSepPunct{\mcitedefaultmidpunct}
{\mcitedefaultendpunct}{\mcitedefaultseppunct}\relax
\EndOfBibitem
\bibitem[May and K{\"{u}}hn(2011)May, and K{\"{u}}hn]{May2011a}
May,~V.; K{\"{u}}hn,~O. \emph{Charg. Energy Transf. Dyn. Mol. Syst. Third Ed.};
  Wiley-VCH Verlag GmbH {\&} Co. KGaA: Weinheim, Germany, 2011\relax
\mciteBstWouldAddEndPuncttrue
\mciteSetBstMidEndSepPunct{\mcitedefaultmidpunct}
{\mcitedefaultendpunct}{\mcitedefaultseppunct}\relax
\EndOfBibitem
\bibitem[Letokhov(1998)]{Mukamel1995}
Letokhov,~V. \emph{Uspekhi Fiz. Nauk}; Oxford University Press, 1998; Vol. 168;
  p 591\relax
\mciteBstWouldAddEndPuncttrue
\mciteSetBstMidEndSepPunct{\mcitedefaultmidpunct}
{\mcitedefaultendpunct}{\mcitedefaultseppunct}\relax
\EndOfBibitem
\bibitem[Duschinsky(1937)]{Duschinsky1937}
Duschinsky,~F. {On the Interpretation of Eletronic Spectra of Polyatomic
  Molecules}. \emph{Acta Physicochim. U.R.S.S.} \textbf{1937}, \emph{7},
  551--566\relax
\mciteBstWouldAddEndPuncttrue
\mciteSetBstMidEndSepPunct{\mcitedefaultmidpunct}
{\mcitedefaultendpunct}{\mcitedefaultseppunct}\relax
\EndOfBibitem
\bibitem[Sando \latin{et~al.}(2001)Sando, Spears, Hupp, and Ruhoff]{Sando2001}
Sando,~G.~M.; Spears,~K.~G.; Hupp,~J.~T.; Ruhoff,~P.~T. {Large electron
  transfer rate effects from the Duschinsky mixing of vibrations}. \emph{J.
  Phys. Chem. A} \textbf{2001}, \emph{105}, 5317--5325\relax
\mciteBstWouldAddEndPuncttrue
\mciteSetBstMidEndSepPunct{\mcitedefaultmidpunct}
{\mcitedefaultendpunct}{\mcitedefaultseppunct}\relax
\EndOfBibitem
\bibitem[Meier and Rauhut(2015)Meier, and Rauhut]{Meier2015}
Meier,~P.; Rauhut,~G. {Comparison of methods for calculating Franck-Condon
  factors beyond the harmonic approximation: How important are Duschinsky
  rotations?} \emph{Mol. Phys.} \textbf{2015}, \emph{113}, 3859--3873\relax
\mciteBstWouldAddEndPuncttrue
\mciteSetBstMidEndSepPunct{\mcitedefaultmidpunct}
{\mcitedefaultendpunct}{\mcitedefaultseppunct}\relax
\EndOfBibitem
\bibitem[Yan and Mukamel(1986)Yan, and Mukamel]{Yan1998}
Yan,~Y.~J.; Mukamel,~S. {Eigenstate-free, Green function, calculation of
  molecular absorption and fluorescence line shapes}. \emph{J. Chem. Phys.}
  \textbf{1986}, \emph{85}, 5908--5923\relax
\mciteBstWouldAddEndPuncttrue
\mciteSetBstMidEndSepPunct{\mcitedefaultmidpunct}
{\mcitedefaultendpunct}{\mcitedefaultseppunct}\relax
\EndOfBibitem
\bibitem[Borrelli and Peluso(2003)Borrelli, and Peluso]{Borrelli2003}
Borrelli,~R.; Peluso,~A. {Dynamics of radiationless transitions in large
  molecular systems: A Franck-Condon-based method accounting for displacements
  and rotations of all the normal coordinates}. \emph{J. Chem. Phys.}
  \textbf{2003}, \emph{119}, 8437--8448\relax
\mciteBstWouldAddEndPuncttrue
\mciteSetBstMidEndSepPunct{\mcitedefaultmidpunct}
{\mcitedefaultendpunct}{\mcitedefaultseppunct}\relax
\EndOfBibitem
\bibitem[Ianconescu and Pollak(2004)Ianconescu, and Pollak]{Ianconescu2004}
Ianconescu,~R.; Pollak,~E. {Photoinduced cooling of polyatomic molecules in an
  electronically excited state in the presence of dushinskii rotations}.
  \emph{J. Phys. Chem. A} \textbf{2004}, \emph{108}, 7778--7784\relax
\mciteBstWouldAddEndPuncttrue
\mciteSetBstMidEndSepPunct{\mcitedefaultmidpunct}
{\mcitedefaultendpunct}{\mcitedefaultseppunct}\relax
\EndOfBibitem
\bibitem[Borrelli \latin{et~al.}(2013)Borrelli, Capobianco, and
  Peluso]{BorrelliRaffaele2013}
Borrelli,~R.; Capobianco,~A.; Peluso,~A. {Franck-Condon factors-Computational
  approaches and recent developments}. \emph{Can. J. Chem.} \textbf{2013},
  \emph{91}, 495--504\relax
\mciteBstWouldAddEndPuncttrue
\mciteSetBstMidEndSepPunct{\mcitedefaultmidpunct}
{\mcitedefaultendpunct}{\mcitedefaultseppunct}\relax
\EndOfBibitem
\bibitem[Niu \latin{et~al.}(2010)Niu, Peng, Deng, Gao, and Shuai]{Niu2010}
Niu,~Y.; Peng,~Q.; Deng,~C.; Gao,~X.; Shuai,~Z. {Theory of excited state decays
  and optical spectra: Application to polyatomic molecules}. \emph{J. Phys.
  Chem. A} \textbf{2010}, \emph{114}, 7817--7831\relax
\mciteBstWouldAddEndPuncttrue
\mciteSetBstMidEndSepPunct{\mcitedefaultmidpunct}
{\mcitedefaultendpunct}{\mcitedefaultseppunct}\relax
\EndOfBibitem
\bibitem[Borrelli \latin{et~al.}(2012)Borrelli, Capobianco, and
  Peluso]{Borrelli2012}
Borrelli,~R.; Capobianco,~A.; Peluso,~A. {Generating function approach to the
  calculation of spectral band shapes of free-base chlorin including Duschinsky
  and Herzberg-Teller effects}. \emph{J. Phys. Chem. A} \textbf{2012},
  \emph{116}, 9934--9940\relax
\mciteBstWouldAddEndPuncttrue
\mciteSetBstMidEndSepPunct{\mcitedefaultmidpunct}
{\mcitedefaultendpunct}{\mcitedefaultseppunct}\relax
\EndOfBibitem
\bibitem[Baiardi \latin{et~al.}(2013)Baiardi, Bloino, and Barone]{Baiardi2013}
Baiardi,~A.; Bloino,~J.; Barone,~V. {General time dependent approach to
  vibronic spectroscopy including franck-condon, herzberg-teller, and
  duschinsky effects}. \emph{J. Chem. Theory Comput.} \textbf{2013}, \emph{9},
  4097--4115\relax
\mciteBstWouldAddEndPuncttrue
\mciteSetBstMidEndSepPunct{\mcitedefaultmidpunct}
{\mcitedefaultendpunct}{\mcitedefaultseppunct}\relax
\EndOfBibitem
\bibitem[Toutounji(2020)]{Toutounji2020}
Toutounji,~M. {Spectroscopy of Vibronically Coupled and Duschinskcally Rotated
  Polyatomic Molecules}. \emph{J. Chem. Theory Comput.} \textbf{2020},
  \emph{16}, 1690--1698\relax
\mciteBstWouldAddEndPuncttrue
\mciteSetBstMidEndSepPunct{\mcitedefaultmidpunct}
{\mcitedefaultendpunct}{\mcitedefaultseppunct}\relax
\EndOfBibitem
\bibitem[Davydov(1979)]{Davydov1979}
Davydov,~A.~S. {Solitons in molecular systems}. \emph{Phys. Scr.}
  \textbf{1979}, \emph{20}, 387--394\relax
\mciteBstWouldAddEndPuncttrue
\mciteSetBstMidEndSepPunct{\mcitedefaultmidpunct}
{\mcitedefaultendpunct}{\mcitedefaultseppunct}\relax
\EndOfBibitem
\bibitem[Scott(1991)]{Scott1991}
Scott,~A.~C. {Davydov's soliton revisited}. \emph{Phys. D Nonlinear Phenom.}
  \textbf{1991}, \emph{51}, 333--342\relax
\mciteBstWouldAddEndPuncttrue
\mciteSetBstMidEndSepPunct{\mcitedefaultmidpunct}
{\mcitedefaultendpunct}{\mcitedefaultseppunct}\relax
\EndOfBibitem
\bibitem[Choi(2004)]{Choi2004}
Choi,~J.~R. {Coherent states of general time-dependent harmonic oscillator}.
  \emph{Pramana - J. Phys.} \textbf{2004}, \emph{62}, 13--29\relax
\mciteBstWouldAddEndPuncttrue
\mciteSetBstMidEndSepPunct{\mcitedefaultmidpunct}
{\mcitedefaultendpunct}{\mcitedefaultseppunct}\relax
\EndOfBibitem
\bibitem[Sun \latin{et~al.}(2010)Sun, Luo, and Zhao]{Sun2010b}
Sun,~J.; Luo,~B.; Zhao,~Y. {Dynamics of a one-dimensional Holstein polaron with
  the Davydov ans{\"{a}}tze}. \emph{Phys. Rev. B - Condens. Matter Mater.
  Phys.} \textbf{2010}, \emph{82}, 014305\relax
\mciteBstWouldAddEndPuncttrue
\mciteSetBstMidEndSepPunct{\mcitedefaultmidpunct}
{\mcitedefaultendpunct}{\mcitedefaultseppunct}\relax
\EndOfBibitem
\bibitem[Choro{\v{s}}ajev \latin{et~al.}(2016)Choro{\v{s}}ajev, Rancova, and
  Abramavicius]{Chorosajev2016a}
Choro{\v{s}}ajev,~V.; Rancova,~O.; Abramavicius,~D. {Polaronic effects at
  finite temperatures in the B850 ring of the LH2 complex}. \emph{Phys. Chem.
  Chem. Phys.} \textbf{2016}, \emph{18}, 7966--7977\relax
\mciteBstWouldAddEndPuncttrue
\mciteSetBstMidEndSepPunct{\mcitedefaultmidpunct}
{\mcitedefaultendpunct}{\mcitedefaultseppunct}\relax
\EndOfBibitem
\bibitem[Wang \latin{et~al.}(2016)Wang, Chen, Zhou, and Zhao]{Wang2016b}
Wang,~L.; Chen,~L.; Zhou,~N.; Zhao,~Y. {Variational dynamics of the sub-Ohmic
  spin-boson model on the basis of multiple Davydov D1 states}. \emph{J. Chem.
  Phys.} \textbf{2016}, \emph{144}, 024101\relax
\mciteBstWouldAddEndPuncttrue
\mciteSetBstMidEndSepPunct{\mcitedefaultmidpunct}
{\mcitedefaultendpunct}{\mcitedefaultseppunct}\relax
\EndOfBibitem
\bibitem[Jaku{\v{c}}ionis \latin{et~al.}(2018)Jaku{\v{c}}ionis,
  Choro{\v{s}}ajev, and Abramavi{\v{c}}ius]{Jakucionis2018c}
Jaku{\v{c}}ionis,~M.; Choro{\v{s}}ajev,~V.; Abramavi{\v{c}}ius,~D. {Vibrational
  damping effects on electronic energy relaxation in molecular aggregates}.
  \emph{Chem. Phys.} \textbf{2018}, \emph{515}, 193--202\relax
\mciteBstWouldAddEndPuncttrue
\mciteSetBstMidEndSepPunct{\mcitedefaultmidpunct}
{\mcitedefaultendpunct}{\mcitedefaultseppunct}\relax
\EndOfBibitem
\bibitem[Jaku{\v{c}}ionis \latin{et~al.}(2020)Jaku{\v{c}}ionis, Mancal, and
  Abramavi{\v{c}}ius]{Jakucionis2020a}
Jaku{\v{c}}ionis,~M.; Mancal,~T.; Abramavi{\v{c}}ius,~D. {Modeling irreversible
  molecular internal conversion using the time-dependent variational approach
  with sD2 ansatz}. \emph{Phys. Chem. Chem. Phys.} \textbf{2020}, \emph{22},
  8952--8962\relax
\mciteBstWouldAddEndPuncttrue
\mciteSetBstMidEndSepPunct{\mcitedefaultmidpunct}
{\mcitedefaultendpunct}{\mcitedefaultseppunct}\relax
\EndOfBibitem
\bibitem[Sun \latin{et~al.}(2015)Sun, Gelin, Chernyak, and Zhao]{Sun2015a}
Sun,~K.~W.; Gelin,~M.~F.; Chernyak,~V.~Y.; Zhao,~Y. {Davydov Ansatz as an
  efficient tool for the simulation of nonlinear optical response of molecular
  aggregates}. \emph{J. Chem. Phys.} \textbf{2015}, \emph{142}, 212448\relax
\mciteBstWouldAddEndPuncttrue
\mciteSetBstMidEndSepPunct{\mcitedefaultmidpunct}
{\mcitedefaultendpunct}{\mcitedefaultseppunct}\relax
\EndOfBibitem
\bibitem[Zhou \latin{et~al.}(2016)Zhou, Chen, Huang, Sun, Tanimura, and
  Zhao]{Zhou2016}
Zhou,~N.; Chen,~L.; Huang,~Z.; Sun,~K.; Tanimura,~Y.; Zhao,~Y. {Fast, Accurate
  Simulation of Polaron Dynamics and Multidimensional Spectroscopy by Multiple
  Davydov Trial States}. \emph{J. Phys. Chem. A} \textbf{2016}, \emph{120},
  1562--1576\relax
\mciteBstWouldAddEndPuncttrue
\mciteSetBstMidEndSepPunct{\mcitedefaultmidpunct}
{\mcitedefaultendpunct}{\mcitedefaultseppunct}\relax
\EndOfBibitem
\bibitem[Choro{\v{s}}ajev \latin{et~al.}(2017)Choro{\v{s}}ajev,
  Mar{\v{c}}iulionis, and Abramavicius]{Chorosajev2017a}
Choro{\v{s}}ajev,~V.; Mar{\v{c}}iulionis,~T.; Abramavicius,~D. {Temporal
  dynamics of excitonic states with nonlinear electron-vibrational coupling}.
  \emph{J. Chem. Phys.} \textbf{2017}, \emph{147}, 74114\relax
\mciteBstWouldAddEndPuncttrue
\mciteSetBstMidEndSepPunct{\mcitedefaultmidpunct}
{\mcitedefaultendpunct}{\mcitedefaultseppunct}\relax
\EndOfBibitem
\bibitem[Somoza \latin{et~al.}(2017)Somoza, Sun, Molina, and Zhao]{Somoza2017a}
Somoza,~A.~D.; Sun,~K.~W.; Molina,~R.~A.; Zhao,~Y. {Dynamics of coherence,
  localization and excitation transfer in disordered nanorings}. \emph{Phys.
  Chem. Chem. Phys.} \textbf{2017}, \emph{19}, 25996--26013\relax
\mciteBstWouldAddEndPuncttrue
\mciteSetBstMidEndSepPunct{\mcitedefaultmidpunct}
{\mcitedefaultendpunct}{\mcitedefaultseppunct}\relax
\EndOfBibitem
\bibitem[Chen \latin{et~al.}(2019)Chen, Gelin, and Domcke]{Chen2019a}
Chen,~L.; Gelin,~M.~F.; Domcke,~W. {Multimode quantum dynamics with multiple
  Davydov D2 trial states: Application to a 24-dimensional conical intersection
  model}. \emph{J. Chem. Phys.} \textbf{2019}, \emph{150}, 24101\relax
\mciteBstWouldAddEndPuncttrue
\mciteSetBstMidEndSepPunct{\mcitedefaultmidpunct}
{\mcitedefaultendpunct}{\mcitedefaultseppunct}\relax
\EndOfBibitem
\bibitem[Glauber(1963)]{Glauber1963}
Glauber,~R.~J. {Coherent and incoherent states of the radiation field}.
  \emph{Phys. Rev.} \textbf{1963}, \emph{131}, 2766--2788\relax
\mciteBstWouldAddEndPuncttrue
\mciteSetBstMidEndSepPunct{\mcitedefaultmidpunct}
{\mcitedefaultendpunct}{\mcitedefaultseppunct}\relax
\EndOfBibitem
\bibitem[Schmidt \latin{et~al.}(1993)Schmidt, Baldridge, Boatz, Elbert, Gordon,
  Jensen, Koseki, Matsunaga, Nguyen, Su, Windus, Dupuis, and
  Montgomery]{Schmidt1993}
Schmidt,~M.~W.; Baldridge,~K.~K.; Boatz,~J.~A.; Elbert,~S.~T.; Gordon,~M.~S.;
  Jensen,~J.~H.; Koseki,~S.; Matsunaga,~N.; Nguyen,~K.~A.; Su,~S.
  \latin{et~al.}  {General atomic and molecular electronic structure system}.
  \emph{J. Comput. Chem.} \textbf{1993}, \emph{14}, 1347--1363\relax
\mciteBstWouldAddEndPuncttrue
\mciteSetBstMidEndSepPunct{\mcitedefaultmidpunct}
{\mcitedefaultendpunct}{\mcitedefaultseppunct}\relax
\EndOfBibitem
\bibitem[Frisch \latin{et~al.}(2016)Frisch, Trucks, Schlegel, Scuseria, Robb,
  Cheeseman, Scalmani, Barone, Petersson, Nakatsuji, Li, Caricato, Marenich,
  Bloino, Janesko, Gomperts, Mennucci, Hratchian, Ortiz, Izmaylov, Sonnenberg,
  Williams-Young, Ding, Lipparini, Egidi, Goings, Peng, Petrone, Henderson,
  Ranasinghe, Zakrzewski, Gao, Rega, Zheng, Liang, Hada, Ehara, Toyota, Fukuda,
  Hasegawa, Ishida, Nakajima, Honda, Kitao, Nakai, Vreven, Throssell,
  Montgomery, Peralta, Ogliaro, Bearpark, Heyd, Brothers, Kudin, Staroverov,
  Keith, Kobayashi, Normand, Raghavachari, Rendell, Burant, Iyengar, Tomasi,
  Cossi, Millam, Klene, Adamo, Cammi, Ochterski, Martin, Morokuma, Farkas,
  Foresman, and Fox]{g16}
Frisch,~M.~J.; Trucks,~G.~W.; Schlegel,~H.~B.; Scuseria,~G.~E.; Robb,~M.~A.;
  Cheeseman,~J.~R.; Scalmani,~G.; Barone,~V.; Petersson,~G.~A.; Nakatsuji,~H.
  \latin{et~al.}  Gaussian 16 {R}evision {C}.01. 2016; Gaussian Inc.
  Wallingford CT\relax
\mciteBstWouldAddEndPuncttrue
\mciteSetBstMidEndSepPunct{\mcitedefaultmidpunct}
{\mcitedefaultendpunct}{\mcitedefaultseppunct}\relax
\EndOfBibitem
\bibitem[Macernis \latin{et~al.}(2014)Macernis, Sulskus, Malickaja, Robert, and
  Valkunas]{Macernis2014}
Macernis,~M.; Sulskus,~J.; Malickaja,~S.; Robert,~B.; Valkunas,~L. {Resonance
  raman spectra and electronic transitions in carotenoids: A density functional
  theory study}. \emph{J. Phys. Chem. A} \textbf{2014}, \emph{118},
  1817--1825\relax
\mciteBstWouldAddEndPuncttrue
\mciteSetBstMidEndSepPunct{\mcitedefaultmidpunct}
{\mcitedefaultendpunct}{\mcitedefaultseppunct}\relax
\EndOfBibitem
\bibitem[Macernis \latin{et~al.}(2015)Macernis, Galzerano, Sulskus, Kish, Kim,
  Koo, Valkunas, and Robert]{Macernis2015}
Macernis,~M.; Galzerano,~D.; Sulskus,~J.; Kish,~E.; Kim,~Y.~H.; Koo,~S.;
  Valkunas,~L.; Robert,~B. {Resonance Raman spectra of carotenoid molecules:
  Influence of methyl substitutions}. \emph{J. Phys. Chem. A} \textbf{2015},
  \emph{119}, 56--66\relax
\mciteBstWouldAddEndPuncttrue
\mciteSetBstMidEndSepPunct{\mcitedefaultmidpunct}
{\mcitedefaultendpunct}{\mcitedefaultseppunct}\relax
\EndOfBibitem
\bibitem[Liu \latin{et~al.}(2010)Liu, Wang, Zheng, Li, and Su]{Wei-Long2010}
Liu,~W.~L.; Wang,~D.~M.; Zheng,~Z.~R.; Li,~A.~H.; Su,~W.~H. {Solvent effects on
  the S0 S2 absorption spectra of $\beta$-carotene}. \emph{Chinese Phys. B}
  \textbf{2010}, \emph{19}, 013102--6\relax
\mciteBstWouldAddEndPuncttrue
\mciteSetBstMidEndSepPunct{\mcitedefaultmidpunct}
{\mcitedefaultendpunct}{\mcitedefaultseppunct}\relax
\EndOfBibitem
\bibitem[Wong(1996)]{Wong1996}
Wong,~M.~W. {Vibrational frequency prediction using density functional theory}.
  \emph{Chem. Phys. Lett.} \textbf{1996}, \emph{256}, 391--399\relax
\mciteBstWouldAddEndPuncttrue
\mciteSetBstMidEndSepPunct{\mcitedefaultmidpunct}
{\mcitedefaultendpunct}{\mcitedefaultseppunct}\relax
\EndOfBibitem
\bibitem[Hirata and Head-Gordon(1999)Hirata, and Head-Gordon]{Hirata1999}
Hirata,~S.; Head-Gordon,~M. {Time-dependent density functional theory within
  the Tamm-Dancoff approximation}. \emph{Chem. Phys. Lett.} \textbf{1999},
  \emph{314}, 291--299\relax
\mciteBstWouldAddEndPuncttrue
\mciteSetBstMidEndSepPunct{\mcitedefaultmidpunct}
{\mcitedefaultendpunct}{\mcitedefaultseppunct}\relax
\EndOfBibitem
\bibitem[Casida(1995)]{Casida1995}
Casida,~M.~E. \emph{{Time-Dependent Density Functional Response Theory for
  Molecules}}; 1995; pp 155--192\relax
\mciteBstWouldAddEndPuncttrue
\mciteSetBstMidEndSepPunct{\mcitedefaultmidpunct}
{\mcitedefaultendpunct}{\mcitedefaultseppunct}\relax
\EndOfBibitem
\bibitem[Dreuw and Head-Gordon(2005)Dreuw, and Head-Gordon]{Dreuw2005}
Dreuw,~A.; Head-Gordon,~M. {Single-reference ab initio methods for the
  calculation of excited states of large molecules}. \emph{Chem. Rev.}
  \textbf{2005}, \emph{105}, 4009--4037\relax
\mciteBstWouldAddEndPuncttrue
\mciteSetBstMidEndSepPunct{\mcitedefaultmidpunct}
{\mcitedefaultendpunct}{\mcitedefaultseppunct}\relax
\EndOfBibitem
\bibitem[Dreuw(2006)]{Dreuw2006}
Dreuw,~A. {Influence of geometry relaxation on the energies of the S1 and S2
  states of violaxanthin, zeaxanthin, and lutein}. \emph{J. Phys. Chem. A}
  \textbf{2006}, \emph{110}, 4592--4599\relax
\mciteBstWouldAddEndPuncttrue
\mciteSetBstMidEndSepPunct{\mcitedefaultmidpunct}
{\mcitedefaultendpunct}{\mcitedefaultseppunct}\relax
\EndOfBibitem
\bibitem[Starcke \latin{et~al.}(2006)Starcke, Wormit, Schirmer, and
  Dreuw]{Starcke2006}
Starcke,~J.~H.; Wormit,~M.; Schirmer,~J.; Dreuw,~A. {How much double excitation
  character do the lowest excited states of linear polyenes have?} \emph{Chem.
  Phys.} \textbf{2006}, \emph{329}, 39--49\relax
\mciteBstWouldAddEndPuncttrue
\mciteSetBstMidEndSepPunct{\mcitedefaultmidpunct}
{\mcitedefaultendpunct}{\mcitedefaultseppunct}\relax
\EndOfBibitem
\bibitem[Balevi{\v{c}}ius \latin{et~al.}(2015)Balevi{\v{c}}ius, Pour,
  Savolainen, Lincoln, Luke{\v{s}}, Riedle, Valkunas, Abramavicius, and
  Hauer]{Balevicius2015}
Balevi{\v{c}}ius,~V.; Pour,~A.~G.; Savolainen,~J.; Lincoln,~C.~N.;
  Luke{\v{s}},~V.; Riedle,~E.; Valkunas,~L.; Abramavicius,~D.; Hauer,~J.
  {Vibronic energy relaxation approach highlighting deactivation pathways in
  carotenoids}. \emph{Phys. Chem. Chem. Phys.} \textbf{2015}, \emph{17},
  19491--19499\relax
\mciteBstWouldAddEndPuncttrue
\mciteSetBstMidEndSepPunct{\mcitedefaultmidpunct}
{\mcitedefaultendpunct}{\mcitedefaultseppunct}\relax
\EndOfBibitem
\bibitem[de~Oliveira \latin{et~al.}(2010)de~Oliveira, Castro, Edwards, and
  de~Oliveiraa]{DeOliveira2009}
de~Oliveira,~V.~E.; Castro,~H.~V.; Edwards,~H.~G.; de~Oliveiraa,~L. F.~C.
  {Carotenes and carotenoids in natural biological samples: A Raman
  spectroscopic analysis}. \emph{J. Raman Spectrosc.} \textbf{2010}, \emph{41},
  642--650\relax
\mciteBstWouldAddEndPuncttrue
\mciteSetBstMidEndSepPunct{\mcitedefaultmidpunct}
{\mcitedefaultendpunct}{\mcitedefaultseppunct}\relax
\EndOfBibitem
\bibitem[Tschirner \latin{et~al.}(2009)Tschirner, Schenderlein, Brose,
  Schlodder, Mroginski, Thomsen, and Hildebrandt]{Tschirner2009a}
Tschirner,~N.; Schenderlein,~M.; Brose,~K.; Schlodder,~E.; Mroginski,~M.~A.;
  Thomsen,~C.; Hildebrandt,~P. {Resonance Raman spectra of $\beta$-carotene in
  solution and in photosystems revisited: an experimental and theoretical
  study}. \emph{Phys. Chem. Chem. Phys.} \textbf{2009}, \emph{11},
  11471--11478\relax
\mciteBstWouldAddEndPuncttrue
\mciteSetBstMidEndSepPunct{\mcitedefaultmidpunct}
{\mcitedefaultendpunct}{\mcitedefaultseppunct}\relax
\EndOfBibitem
\bibitem[Pol{\'{i}}vka and Sundstr{\"{o}}m(2009)Pol{\'{i}}vka, and
  Sundstr{\"{o}}m]{Polivka2009}
Pol{\'{i}}vka,~T.; Sundstr{\"{o}}m,~V. {Dark excited states of carotenoids:
  Consensus and controversy}. \emph{Chem. Phys. Lett.} \textbf{2009},
  \emph{477}, 1--11\relax
\mciteBstWouldAddEndPuncttrue
\mciteSetBstMidEndSepPunct{\mcitedefaultmidpunct}
{\mcitedefaultendpunct}{\mcitedefaultseppunct}\relax
\EndOfBibitem
\bibitem[Roscioli \latin{et~al.}(2017)Roscioli, Ghosh, LaFountain, Frank, and
  Beck]{Roscioli2017}
Roscioli,~J.~D.; Ghosh,~S.; LaFountain,~A.~M.; Frank,~H.~A.; Beck,~W.~F.
  {Quantum Coherent Excitation Energy Transfer by Carotenoids in Photosynthetic
  Light Harvesting}. \emph{J. Phys. Chem. Lett.} \textbf{2017}, \emph{8},
  5141--5147\relax
\mciteBstWouldAddEndPuncttrue
\mciteSetBstMidEndSepPunct{\mcitedefaultmidpunct}
{\mcitedefaultendpunct}{\mcitedefaultseppunct}\relax
\EndOfBibitem
\bibitem[Meneghin \latin{et~al.}(2018)Meneghin, Volpato, Cupellini, Bolzonello,
  Jurinovich, Mascoli, Carbonera, Mennucci, and Collini]{Meneghin2018a}
Meneghin,~E.; Volpato,~A.; Cupellini,~L.; Bolzonello,~L.; Jurinovich,~S.;
  Mascoli,~V.; Carbonera,~D.; Mennucci,~B.; Collini,~E. {Coherence in
  carotenoid-to-chlorophyll energy transfer}. \emph{Nat. Commun.}
  \textbf{2018}, \emph{9}, 3160\relax
\mciteBstWouldAddEndPuncttrue
\mciteSetBstMidEndSepPunct{\mcitedefaultmidpunct}
{\mcitedefaultendpunct}{\mcitedefaultseppunct}\relax
\EndOfBibitem
\bibitem[Ghosh \latin{et~al.}(2017)Ghosh, Bishop, Roscioli, LaFountain, Frank,
  and Beck]{Ghosh2017}
Ghosh,~S.; Bishop,~M.~M.; Roscioli,~J.~D.; LaFountain,~A.~M.; Frank,~H.~A.;
  Beck,~W.~F. {Excitation Energy Transfer by Coherent and Incoherent Mechanisms
  in the Peridinin-Chlorophyll a Protein}. \emph{J. Phys. Chem. Lett.}
  \textbf{2017}, \emph{8}, 463--469\relax
\mciteBstWouldAddEndPuncttrue
\mciteSetBstMidEndSepPunct{\mcitedefaultmidpunct}
{\mcitedefaultendpunct}{\mcitedefaultseppunct}\relax
\EndOfBibitem
\bibitem[Perl{\'{i}}k \latin{et~al.}(2015)Perl{\'{i}}k, Seibt, Cranston,
  Cogdell, Lincoln, Savolainen, {\v{S}}anda, Man{\v{c}}al, and
  Hauer]{Perlik2015b}
Perl{\'{i}}k,~V.; Seibt,~J.; Cranston,~L.~J.; Cogdell,~R.~J.; Lincoln,~C.~N.;
  Savolainen,~J.; {\v{S}}anda,~F.; Man{\v{c}}al,~T.; Hauer,~J. {Vibronic
  coupling explains the ultrafast carotenoid-to-bacteriochlorophyll energy
  transfer in natural and artificial light harvesters}. \emph{J. Chem. Phys.}
  \textbf{2015}, \emph{142}, 212434\relax
\mciteBstWouldAddEndPuncttrue
\mciteSetBstMidEndSepPunct{\mcitedefaultmidpunct}
{\mcitedefaultendpunct}{\mcitedefaultseppunct}\relax
\EndOfBibitem
\end{mcitethebibliography}
\label{reference_label}\newpage\includegraphics[width=8.25cm]{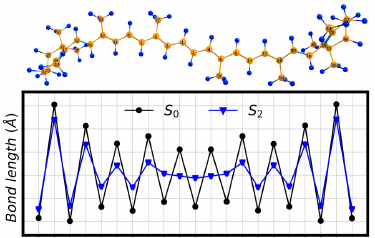}\label{TOC Graphic}
\end{document}